\pdfoutput=1
\documentclass[conference]{IEEEtran}

\usepackage[utf8]{inputenc}
\usepackage{microtype}
\usepackage{graphicx}
\usepackage{booktabs}
\usepackage{array}
\newcolumntype{L}[1]{>{\raggedright\arraybackslash}p{#1}}
\usepackage{multirow}
\usepackage{amsmath}
\usepackage{url}
\usepackage{tikz}
\usepackage{enumitem}
\usepackage{fontawesome5}
\usetikzlibrary{shapes.geometric,arrows.meta,positioning}
\usepackage{xcolor}
\usepackage[ruled,vlined]{algorithm2e}
\usepackage{cite}
\usepackage[font=footnotesize,labelfont=bf,labelsep=period,justification=justified,singlelinecheck=true]{caption}

\usepackage[most]{tcolorbox}
\newtcolorbox{answerbox}[1]{enhanced, breakable, colback=black!4,
  colframe=black!55, boxrule=0.5pt, arc=1.5pt, left=4pt, right=4pt,
  top=2pt, bottom=2pt, fonttitle=\bfseries\footnotesize, coltitle=white,
  colbacktitle=black!65, title={#1}, fontupper=\footnotesize}
\usepackage[colorlinks=true,linkcolor=blue!60!black,citecolor=green!40!black,urlcolor=blue!60!black]{hyperref}
\hypersetup{
  pdftitle={Noise-Awareness in Quantum Software Testing: An Empirical Landscape Analysis},
  pdfauthor={Mir Mohammad Yousuf, Shabir Ahmad Sofi, Bisma Majid},
  pdfkeywords={Quantum software testing, NISQ, noise-aware testing, systematic mapping study, empirical software engineering}
}

\begin{document}

\title{Noise-Awareness in Quantum Software Testing:\\
An Empirical Landscape Analysis}

\author{
\IEEEauthorblockN{Mir Mohammad Yousuf}
\IEEEauthorblockA{\textit{Department of Information Technology}\\
\textit{National Institute of Technology}\\
Srinagar, J\&K, India\\
yousuf\_2022phaite006@nitsri.ac.in\\
\textit{Corresponding author}}
\and
\IEEEauthorblockN{Shabir Ahmad Sofi}
\IEEEauthorblockA{\textit{Department of Information Technology}\\
\textit{National Institute of Technology}\\
Srinagar, J\&K, India\\
shabir@nitsri.ac.in}
\and
\IEEEauthorblockN{Bisma Majid}
\IEEEauthorblockA{\textit{Department of Information Technology}\\
\textit{National Institute of Technology}\\
Srinagar, J\&K, India\\
bismabhat\_ite006@nitsri.ac.in}
}

\maketitle
\bstctlcite{IEEEexample:BSTcontrol}

\begin{abstract}
Noisy Intermediate-Scale Quantum (NISQ) hardware has transformed quantum computing into a practical experimental platform while introducing fundamental challenges associated with hardware noise, decoherence, and probabilistic execution. Although quantum software testing (QST) has rapidly emerged as an active research area, it remains unclear to what extent existing testing methodologies explicitly engage with noisy and hardware-constrained execution environments. This paper presents a systematic empirical landscape analysis of noise-awareness in QST research through a deterministic, taxonomy-guided analytical framework applied to 61 papers published between 2018 and 2025 across eight testing paradigms, contrasting how contributions are framed with how they are evaluated. Among the analyzed papers, 47 (77.0\%) are noise-oblivious and only 14 (23.0\%) explicitly engage with noise-related execution concerns, while only 11 (18.0\%) report evaluation on physical hardware or a noisy simulator. Test generation, the largest paradigm, exhibits a noise-awareness rate of only 14.3\%, and debugging and repair 18.2\%; coverage-based, metamorphic, and differential testing contain no noise-aware contributions. Noise-aware methodologies emerged late and remain outnumbered in all analyzed years. Framing and evaluation are strongly associated (odds ratio 40.5, $p < 0.001$) yet diverge for individual papers: 5 of the 14 noise-aware papers evaluate only on ideal simulators. Inferential statistical tests are reported in only 16.4\% of papers. The classification framework was validated against a blinded, human-coded reference sample, achieving almost perfect inter-rater agreement on testing-technique classification and substantial agreement on noise-awareness. The framework provides a reusable empirical basis for longitudinal analysis of hardware-conscious testing practice.
\end{abstract}


\section{Introduction}
\label{sec:introduction}

Quantum computing has progressed from a theoretical paradigm of algorithmic possibility into a tangible experimental discipline supported by physical hardware. Contemporary platforms based on superconducting qubits, trapped ions, neutral atoms, and photonic architectures now enable the routine execution of nontrivial quantum programs at scales of tens to hundreds of qubits~\cite{preskill2018nisq, krantz2019quantum, yousuf2026systematic}. This transition is widely characterized as the \emph{Noisy Intermediate-Scale Quantum (NISQ)} era: a period in which quantum devices are sufficiently capable to execute meaningful computations yet remain subject to substantial hardware imperfections, including decoherence, gate errors, measurement noise, crosstalk, and calibration drift~\cite{nielsen2010quantum}. Unlike future fault-tolerant systems, NISQ hardware provides no architectural insulation between algorithmic specification and physical-layer noise; consequently, the observed behavior of a quantum program is shaped jointly by its logical structure and the stochastic dynamics of the underlying device.

This dual dependence introduces a class of engineering challenges that has no direct analogue in classical software. Quantum programs execute under \emph{probabilistic semantics}: identical inputs produce distributions over outcomes rather than deterministic results. They are subject to \emph{hardware-induced uncertainty} that varies across backends, qubit topologies, and even time-of-day calibration profiles. They depend on \emph{transpilation pipelines} whose decisions about gate decomposition and qubit mapping can materially alter observable behavior. They typically lack \emph{ground-truth oracles} suitable for direct correctness assertions, particularly for algorithms whose output distributions cannot be efficiently classically simulated. And they exhibit a persistent \emph{simulator-versus-hardware gap}: a program that passes all simulator-based tests may fail systematically on physical devices, while a program that performs acceptably on one backend may behave very differently on another. These properties collectively make quantum software validation a research problem in its own right, rather than a straightforward extension of classical software testing.

The emerging discipline of \emph{quantum software engineering} (QSE), and particularly the subarea of \emph{quantum software testing} (QST), has responded to these challenges by adapting and extending classical testing paradigms. Recent years have seen the introduction of mutation testing frameworks for quantum programs~\cite{fortunato2022qmutpy}, metamorphic testing approaches that exploit invariances of quantum circuits~\cite{paltenghi2023morphq}, differential testing infrastructures comparing quantum software stacks~\cite{wang2021qdiff}, coverage-guided test generation tools~\cite{wang2021quito}, and a growing body of work on assertions, unit testing of quantum subroutines, debugging, and program repair. Surveys, methodological reviews, and community roadmaps have begun to consolidate this literature~\cite{garcia2023quantum, li2026methodological, murillo2025quantum}, signalling the maturation of QST as a recognizable research subfield.

Despite this progress, a fundamental question concerning the methodological alignment of QST research with NISQ realities remains unanswered: \emph{to what extent does existing QST research explicitly account for hardware noise and realistic NISQ execution environments?} Available surveys characterize the diversity of testing techniques but do not systematically quantify whether proposed methodologies engage with noise as a first-class concern versus assume idealized or noise-suppressed execution. Existing empirical studies of QST tools and practices likewise emphasize methodological breadth over the noise-aware dimension. This gap is not merely descriptive: the answer determines whether the current QST literature provides foundations adequate for testing programs that must execute on real NISQ devices, or whether prevailing methodologies remain anchored to assumptions inherited from classical, deterministic software.

This paper addresses that question through a systematic empirical landscape analysis of 61 QST research contributions published between 2018 and 2025. Central to the study is a deterministic, ambiguity-aware analytical framework designed to characterize how QST methodologies engage with noisy and hardware-constrained execution environments. The framework combines taxonomy-guided classification, semantic-overlap handling, confidence estimation, structured review integration, and an operationalized notion of noise-awareness based on explicit noise indicators, hardware-context indicators, and idealization markers. Together, these components support reproducible and auditable analysis of the methodological alignment between contemporary QST research and NISQ-era execution realities.

Using this analytical framework, we organize the literature along eight testing paradigms, quantify noise-awareness rates and evaluation-realism patterns within and across paradigms, characterize the temporal evolution of both dimensions, and synthesize the results into a volume-versus-awareness view of the QST landscape that exposes structural gaps in the field's current methodological coverage.

Our analysis surfaces four findings of direct relevance to the design of NISQ-aware testing methodologies. First, 77.0\% of the analyzed contributions remain noise-oblivious under the adopted operationalization. Second, the most extensively studied paradigms (test generation and debugging/repair) exhibit noise-awareness rates below the corpus mean, while coverage-based testing (0/5), metamorphic testing (0/2), and differential testing (0/2) contain no noise-aware contributions. Third, unit testing of quantum subroutines exhibits the highest noise-awareness proportion (3 of 6, tied with the residual \emph{Other} category), tentatively suggesting that subroutine-level testing offers a more tractable setting for hardware-aware design, although the difference from test generation is not statistically significant at this sample size ($p \approx 0.13$). Fourth, the methodological elements most directly required by probabilistic NISQ execution (explicit oracles, shot budgets, and inferential statistical evaluation) are also the least consistently reported, even within the noise-aware subset.

The contributions of this paper are summarized as follows:
\begin{enumerate}

  \item A deterministic, ambiguity-aware classification framework for QST literature that combines taxonomy-guided categorization, semantic-overlap handling, confidence estimation, structured review integration, and an operationalized notion of noise-awareness. The framework is designed to be reapplied unchanged to future publication cohorts.

  \item The first systematic empirical landscape analysis of noise-awareness in QST, covering 61 contributions across eight primary paradigms (plus a residual \emph{Other} category) from 2018 to 2025, characterizing the distribution, depth, and temporal evolution of noise-awareness and synthesizing them into a volume-versus-awareness view of the field.

  \item A two-dimensional evidence base that crosses lexical framing with reported evaluation realism and an empirical-practice-pattern aggregation (oracles, statistical evaluation, shot budgets, benchmarks, and reproducibility artifacts), showing that framing and evaluation are strongly associated overall while diverging for individual papers.

  \item A research agenda for NISQ-aware QST that maps each identified gap to a concrete direction, including probabilistic oracles and adequacy criteria, noise-aware relational and differential testing, and cross-platform benchmarking.

\end{enumerate}

The remainder of this paper is organized as follows. Section~\ref{sec:background} provides background on NISQ computing and the testing challenges it raises. Section~\ref{sec:related} reviews related work. Section~\ref{sec:methodology} describes the research design, dataset, classification framework, and validation procedures. Section~\ref{sec:empirical-landscape} reports the findings, organized by research question. Section~\ref{sec:discussion} discusses implications and a research agenda, Section~\ref{sec:threats} addresses threats to validity, and Section~\ref{sec:conclusion} concludes.

\section{Background}
\label{sec:background}

This section establishes the technical context within which our empirical analysis is situated. We summarize the engineering characteristics of NISQ-era quantum computing, the nature of quantum noise as it manifests in observable program behavior, the structural features of hybrid quantum--classical systems, and the resulting challenges that distinguish quantum software testing from its classical counterpart. The discussion is intentionally compact and oriented toward QST researchers; we refer the reader to recent surveys for broader coverage of quantum software engineering~\cite{garcia2023quantum, yousuf2026systematic}.

\subsection{NISQ Computing and Engineering Constraints}
\label{subsec:bg_nisq}

The NISQ regime~\cite{preskill2018nisq} is characterized by quantum processors with sufficient qubits to perform computations beyond the reach of brute-force classical simulation, but without the overhead-intensive quantum error correction required for fault-tolerant operation. Contemporary NISQ devices exhibit several engineering properties that directly shape the behavior and testability of quantum software: limited coherence times that constrain effective circuit depth, gate fidelities that decline multiplicatively with circuit complexity, qubit-connectivity constraints that necessitate SWAP-heavy compilation for non-adjacent two-qubit operations, and time-varying calibration profiles that introduce non-stationary error characteristics. The cumulative effect is that the same logical quantum program may produce materially different output distributions across backends, across calibration cycles, and across compilation strategies.

These constraints are not transient implementation details: they define the operating envelope within which quantum software must currently be engineered. Practical NISQ programming therefore involves explicit reasoning about hardware characteristics (qubit selection, gate scheduling, mitigation strategies, and execution shot budgets) rather than purely algorithmic specification. Any testing methodology intended to validate quantum software for NISQ execution must accommodate this hardware-software coupling~\cite{murillo2025quantum}.

\subsection{Quantum Noise and Probabilistic Execution}
\label{subsec:bg_noise}

Quantum noise refers to the collection of physical mechanisms that perturb the ideal evolution of a quantum state during program execution. Principal sources include decoherence (the loss of quantum coherence through environmental coupling, parameterized by T1 relaxation and T2 dephasing times), gate errors (deviations of applied unitaries from their specifications), readout errors (measurement misclassification), and crosstalk (unintended coupling between qubits during concurrent operations)~\cite{nielsen2010quantum}. For QST purposes, the operationally relevant consequence of these mechanisms is that quantum program execution becomes \emph{probabilistic in a non-trivial sense}: rather than producing a single result per run, a program produces a distribution over results that depends jointly on the program's logical structure, the device's noise profile, and the chosen number of execution shots.

This probabilistic regime complicates several foundational concepts of software testing. The notion of a deterministic correctness oracle becomes ill-posed when the expected behavior is itself a distribution. Test pass/fail decisions require statistical reasoning over sampled outcomes rather than direct comparison of return values. Test flakiness, traditionally treated as a defect of test design in classical software, becomes an intrinsic feature of hardware execution. And the distinction between a genuine logical defect and a noise-induced deviation cannot generally be made from observation of a single execution.

\subsection{Hybrid Execution and Transpilation}
\label{subsec:bg_hybrid}

Most practical quantum software, including the dominant family of variational quantum algorithms~\cite{cerezo2021variational}, executes as a hybrid pipeline in which a classical controller orchestrates parameterized quantum circuits, ingests measurement outcomes, and dispatches updated parameters back to the quantum backend. Defects may therefore manifest in the classical portion, the quantum portion, or the interface between them, and may be reproducible (classical logic), partially reproducible (deterministic transpilation but noisy execution), or non-reproducible across hardware-execution cycles. Before execution, the transpilation pipeline decomposes abstract gates into native ones, maps logical qubits onto physical qubits subject to connectivity constraints, and optimizes circuit depth and gate count~\cite{murali2019noise}. Transpilation is generally optimization-dependent, so equivalent source programs may yield different physical circuits, and it interacts with noise: a deeper or more SWAP-laden compiled circuit accumulates more hardware error even when the source program is unchanged. A testing methodology that evaluates programs only prior to transpilation, or only on a single transpilation result, may therefore miss defects that manifest in the hardware-executable form.

\subsection{Quantum Software Testing under NISQ Conditions}
\label{subsec:bg_qst_under_nisq}

The combination of probabilistic execution, hardware heterogeneity, hybrid pipelines, and transpilation variability gives rise to a set of testing challenges that are characteristic of (though not exclusive to) the quantum domain~\cite{murillo2025quantum}. We summarize four that recur across the QST literature.

\paragraph*{Oracle construction} The classical software-testing literature has long recognized oracle construction as a foundational challenge~\cite{barr2014oracle}. In the quantum domain, this challenge is amplified: for many quantum programs of practical interest, no efficient classical reference implementation is available. Even when a reference is available (e.g., a smaller simulable instance), translating it into a usable oracle requires statistical comparison of distributions rather than equality checking, and demands principled choices of sample size, significance thresholds, and tolerance to noise-induced deviation.

\paragraph*{Test adequacy under stochastic execution} Classical adequacy criteria such as branch or path coverage assume deterministic control flow. Quantum programs exhibit measurement-induced branching and probabilistic state collapse, which require adequacy notions defined over probability distributions or information-theoretic exposure rather than control-flow events.

\paragraph*{Failure localization} When a quantum test fails, the underlying cause may lie in the source program, the transpilation, the hardware execution, or the statistical interpretation. Existing debugging tools rarely span this full causal chain, and most assume that observed deviations can be attributed to a single layer.

\paragraph*{Simulator-versus-hardware validation} Simulator-based evaluation, while indispensable, cannot in general substitute for hardware-based evaluation, because simulators model only an approximation of the noise environment of any specific device. QST methodologies that validate exclusively against simulators are therefore vulnerable to systematic blind spots when applied to real NISQ executions.

These four challenges motivate the central question of the present study: not whether QST exists as a research discipline (it manifestly does) but whether its methodological output explicitly engages with the NISQ realities that practical quantum software must operate within. The remainder of this paper investigates this question empirically.

\section{Related Work}
\label{sec:related}

\paragraph*{Testing techniques for quantum programs} Most QST techniques adapt established classical paradigms, whose foundations include mutation analysis~\cite{jia2010analysis}, the test-oracle problem~\cite{barr2014oracle}, metamorphic testing~\cite{segura2016survey}, and differential testing~\cite{mckeeman1998differential}. Quantum counterparts now exist for each: mutation-analysis tools such as \emph{QMutPy}~\cite{fortunato2022qmutpy} and \emph{Muskit}~\cite{paper_1}; metamorphic testing of quantum computing platforms~\cite{paltenghi2023morphq} and of oracle quantum programs~\cite{abreu2022metamorphic}; differential testing of quantum software stacks with \emph{QDiff}~\cite{wang2021qdiff}; and test generation through coverage guidance (\emph{Quito}~\cite{wang2021quito}), search (\emph{QuSBT}~\cite{paper_17}), combinatorial testing (\emph{QuCAT}~\cite{paper_5}), fuzzing (\emph{QuanFuzz}~\cite{paper_103}), and concolic execution~\cite{paper_45}. A parallel line addresses correctness checking directly through statistical assertions~\cite{huang2019statistical}, projection-based assertions~\cite{li2020projection}, dynamic runtime assertions built from quantum circuits~\cite{paper_122}, statistical testing via fixed-point amplitude amplification~\cite{kang2024statistical}, and design by contract~\cite{yamaguchi2025practical}. Debugging and repair have followed, from general debugging frameworks~\cite{paper_54} to automated program repair~\cite{paper_44}, as has unit testing of quantum subroutines~\cite{paper_114}. These contributions establish \emph{what} can be tested in quantum programs; the present study asks under which execution assumptions they do so.

\paragraph*{Noise in quantum software testing} Hardware noise has entered QST along several routes. Some studies characterize its effect on program behavior~\cite{paper_67} or separate software bugs from hardware noise statistically~\cite{paper_116}; others adapt testing to noisy execution, for example through machine-learning-based noise mitigation in quantum software testing~\cite{paper_73}, property-based testing adapted to real-world noisy computers~\cite{paper_94}, or debugging assertions evaluated under noisy simulation~\cite{paper_115}. A smaller group validates techniques on physical devices, including online monitoring through test points~\cite{paper_15} and testing through commuting Pauli strings~\cite{paper_39} on IBM hardware. From the systems side, noise-adaptive compiler mappings~\cite{murali2019noise} adapt compilation itself to device noise. These works demonstrate that noise-aware testing is feasible, but each addresses noise within a single technique; to our knowledge, none measures how far such engagement extends across the QST literature.

\paragraph*{Secondary studies} García de la Barrera et al.~\cite{garcia2023quantum} review the state of the art in quantum software testing, and Paltenghi and Pradel~\cite{paltenghi2024survey} catalogue 102 works published between 2007 and 2023 across testing, bug-pattern detection, and circuit-analysis dimensions, emphasizing taxonomy and thematic organization rather than quantitative measurement along a single methodological axis. Roadmaps and systematic reviews of quantum software engineering~\cite{murillo2025quantum, yousuf2026systematic} chart open challenges for the broader discipline. Closest in method, Li et al.~\cite{li2026methodological} analyze how empirical QST studies are designed, conducted, and reported, profiling programs under test, execution backends, statistical repetitions, and tooling. To our knowledge, none of these studies operationalizes noise-awareness as a measurable property of QST contributions.

\paragraph*{Positioning} The present study differs from prior work in three respects. It defines noise-awareness as an explicit, reproducible classification dimension and measures it across all major QST paradigms; it crosses that dimension with independently extracted evaluation evidence, which allows framing and validation practice to be compared paper by paper; and it validates the resulting classifications against blinded human coding. The study builds on the primary-study set and evaluation schema of~\cite{li2026methodological} (Section~\ref{subsec:dataset_construction}), while the taxonomy, noise-awareness operationalization, cross-dimensional analyses, and validation are original to this work.

\section{Methodology}\label{sec:methodology}

We conducted a systematic empirical landscape analysis of noise-awareness in quantum software testing (QST) research, following established guidelines for systematic mapping studies in empirical software engineering~\cite{keele2007guidelines, petersen2008systematic}. The methodology combines three components: a taxonomy-guided lexical classification of QST contributions; a structured manual review of cases flagged as ambiguous by that classification; and a complementary analysis of evaluation realism based on reported experimental configurations. The taxonomy was additionally validated through a blinded inter-rater agreement study (Section~\ref{subsubsec:interrater_validation}).

\subsection{Research Design}
\label{subsec:research_design}

The study is designed as an observational, longitudinal mapping analysis of the QST research literature. We chose a mapping-study orientation rather than a narrative review or meta-analysis for two reasons. First, the heterogeneity of QST methodologies, spanning test generation, oracle construction, mutation analysis, metamorphic relations, differential testing, debugging, and quantum-specific adequacy criteria, precludes aggregation of effect sizes across studies. Second, our central research questions concern the \emph{distribution and evolution} of methodological practices across the literature rather than the comparative effectiveness of any particular technique. A rule-based mapping study is therefore the appropriate instrument.

The empirical analysis is guided by four research questions that progressively characterize the state of noise-awareness in QST research:

\begin{itemize}
  \item \textbf{RQ1.} How prevalent is noise-awareness in QST research?
  \item \textbf{RQ2.} How does noise-awareness vary across testing paradigms?
  \item \textbf{RQ3.} How has noise-awareness evolved over time?
  \item \textbf{RQ4.} What major methodological and research gaps remain in noise-aware QST?
\end{itemize}

The unit of analysis is the individual research publication. Classification is performed along two orthogonal dimensions (testing technique and noise-awareness), which are determined independently using the operational definitions developed in Sections~\ref{subsec:classification_framework} and~\ref{subsec:noise_operationalization}.

\subsection{Dataset Construction}
\label{subsec:dataset_construction}

The analyzed dataset comprises 61 QST publications appearing between 2018 and 2025. The publication window was selected to span the emergence of the modern QST literature and to ensure that each annual cohort represents a complete calendar year of research activity; partial-year data from 2026 was excluded to avoid introducing temporal artifacts into the longitudinal analysis.

The corpus was initialized from the 59 primary studies identified in the methodological analysis of empirical QST research by Li et al.~\cite{li2026methodological}, which were collected through keyword search across IEEE Xplore, the ACM Digital Library, Web of Science, and Wiley, complemented by forward and backward snowballing, in line with established guidelines for systematic literature studies~\cite{keele2007guidelines, petersen2008systematic}. We extended this set through a targeted search applying the same inclusion criteria. Search queries combined QST-related terms (e.g., \emph{quantum test*}, \emph{quantum software testing}, \emph{quantum program testing}) with broader validation terminology. To be retained, a paper had to satisfy three inclusion criteria: (i)~it proposes, evaluates, or analyzes a testing-related methodology for quantum programs; (ii)~it appeared between 2018 and 2025; and (iii)~it is peer-reviewed or is available as a preprint with sufficient methodological detail to support classification. Two additional eligible studies were identified and manually verified, yielding a final corpus of 61 papers.

For each retained publication, we extracted the title, abstract, and BibTeX-level keywords (where available) as the textual basis for classification. We deliberately confined the analyzed corpus to metadata rather than full text to ensure that the classification framework operates over information that is uniformly available across papers and reproducibly accessible by independent investigators. Evaluation-realism attributes (Section~\ref{subsec:evaluation_extraction}) are obtained through a separate protocol described in that section.

\subsection{Deterministic Taxonomy-Guided Classification Framework}
\label{subsec:classification_framework}

Classification along the testing-technique dimension is performed through a deterministic, rule-based pipeline. This choice operationalizes methodological identity in a form that is reproducible, auditable, and stable under the sparse metadata conditions in which the classifier operates.

\subsubsection{Testing Technique Taxonomy}
\label{subsubsec:taxonomy}

We organize testing approaches into eight primary categories: \emph{Mutation Testing}; \emph{Oracle/Assertion-Based Testing}; \emph{Metamorphic Testing}; \emph{Differential Testing}; \emph{Unit Testing}; \emph{Coverage-Based Testing}; \emph{Test Generation}; and \emph{Debugging, Repair, and Fault Localization}. Operational definitions follow established surveys and foundational work~\cite{jia2010analysis, barr2014oracle, segura2016survey, mckeeman1998differential}.

The taxonomy extends the canonical six-category structure of those surveys by promoting Unit Testing and Coverage-Based Testing to first-class categories. This expansion responds to a pattern in the analyzed corpus: a non-trivial subset of recent contributions explicitly frames itself around testing of quantum subroutines or around coverage-based adequacy criteria, with dedicated tools and adequacy notions distinct from generic test generation. Treating these as separate categories preserves the resolution at which emerging QST paradigms can be characterized.

Plausible alternatives include a conservative six-category taxonomy that subsumes Unit Testing and Coverage-Based Testing under Test Generation, a fault-oriented grouping of Mutation Testing with Debugging/Repair, and a finer disaggregation of the residual Other category. Such alternatives would change per-paradigm percentages but not the corpus-wide rates of noise-awareness (23.0\%) or hardware-realistic evaluation (18.0\%), which are computed over the full dataset. We retain the eight-category taxonomy because separating Unit Testing from Test Generation exposes a substantially higher noise-awareness rate in the former that a coarser taxonomy would obscure; a robustness check against the six-category alternative is reported in Section~\ref{subsec:threats_taxonomy}.

The taxonomy is intentionally \emph{methodology-oriented} rather than \emph{artifact-oriented}. Papers are classified by the dominant testing mechanism that the proposed contribution embodies, not by secondary mechanisms that appear in implementation details. For example, a metamorphic testing framework that employs assertions for relational validation is classified under Metamorphic Testing rather than Oracle/Assertion-Based Testing; similarly, a behavioral-equivalence framework that uses generated test cases for validation is classified under Differential Testing rather than Test Generation. This design choice reduces over-classification of general-purpose mechanisms (assertions, test generation, debugging) and preserves taxonomic precision in the face of contributions that mix multiple testing concepts.

The complete keyword framework is summarized in Table~\ref{tab:testing_keywords}; no keyword was added or removed after the pipeline produced its final results. Matching uses precompiled word-boundary regular expressions over the lowercased concatenation of title, abstract, and BibTeX keywords, which prevents spurious substring matches (e.g., \emph{mutation} within \emph{permutation}).

\begin{table*}[!t]
\caption{Keyword framework used for classifying testing techniques. Keywords were derived from established terminology in classical and quantum software testing literature and refined through iterative ambiguity analysis on a held-out subset of the corpus. Matching is performed case-insensitively against the unified textual representation of each paper using word-boundary regular expressions.}
\label{tab:testing_keywords}
\centering
\footnotesize
\begin{tabular}{L{4cm} L{11cm}}
\toprule
\textbf{Testing Category} & \textbf{Representative Keywords} \\
\midrule
Mutation Testing & mutation, mutant, mutator, mutation operator, mutation testing, mutation analysis, mutated \\
Oracle / Assertion-Based Testing & test oracle, oracle, oracles, oracle construction, oracle-based testing, assertion, assertions, contract, invariant, property checking, expected output, postcondition, precondition, online monitoring, monitoring, runtime monitoring, specification-based testing, runtime verification, statistical testing \\
Metamorphic Testing & metamorphic, metamorphic relation, metamorphic property, follow-up test, source test case, follow-up test case \\
Differential Testing & differential, cross-checking, reference implementation, comparison testing, equivalence checking, simulator vs hardware, cross-platform testing, reference oracle, n-version, behavioral equivalence, unitarity checking, identity checking, black-box testing \\
Unit Testing & unit testing, unit test, unit tests, quantum unit testing, subroutine testing, testing subroutines, integration testing, quantum subroutine, multiple subroutines, context-aware unit testing \\
Coverage-Based Testing & coverage-guided, coverage criteria, coverage metric, coverage metrics, branch coverage, gate coverage, input coverage, output coverage, coverage testing, coverage analysis \\
Test Generation & test generation, test case generation, automated test generation, search-based testing, fuzzing, fuzz testing, combinatorial testing, random testing, symbolic execution, concolic testing, constraint solving, genetic algorithm, property-based testing, quickcheck, test input generation, test suite generation, test pattern generation, input generation \\
Debugging / Repair / Fault Localization & debugging, debug, debugger, repair, fault localization, program repair, bug fix, diagnosis, spectrum-based fault localization, SBFL, delta debugging, automated debugging, root cause analysis, defect prediction, program slicing, bug locating, bug-locating, locating bug \\
\bottomrule
\end{tabular}
\end{table*}

\subsubsection{Hierarchy-Aware Assignment and Tie-Breaking}
\label{subsubsec:tie_breaking}

Categories are assigned through a hierarchy-aware matching strategy. Because titles and abstracts often carry overlapping or incomplete methodological signals, the rules below act as heuristics that produce a single deterministic outcome under sparse metadata while routing ambiguous cases to manual review. Five rules govern the assignment.

\begin{enumerate}[label=(\roman*)]
  \item Title-level keyword matches receive higher weight than abstract-level matches, on the assumption that titles are authored to communicate the central methodological focus of a paper.

  \item When the title matches keywords from more than one category, the paper is flagged for manual review (\emph{multiple-title-match}) rather than auto-resolved by priority alone. The flagging mechanism is described in Section~\ref{subsec:reliability_validation}.

  \item A semantic override applies when a title contains both debugging-oriented keywords (e.g., \emph{debugging}, \emph{fault localization}) and oracle- or assertion-oriented keywords (e.g., \emph{assertion}, \emph{runtime verification}). The paper is provisionally assigned to Debugging/Repair on the rationale that assertion mechanisms in such titles typically support a debugging purpose, and is flagged for manual review (\emph{oracle+debugging-overlap}).

  \item In the absence of title-level evidence, the category with the highest keyword frequency in the unified corpus is selected, provided the maximum frequency meets a minimum-evidence threshold of $\tau = 2$ occurrences. The threshold is set above singleton occurrence to suppress spurious assignments from isolated or contextually ambiguous terminology.

  \item Ties at the maximum-frequency level are resolved using a priority hierarchy:\\ \emph{Mutation} $>$ \emph{Metamorphic} $>$ \emph{Differential} $>$ \emph{Oracle/Assertion} $>$ \emph{Unit Testing} $>$ \emph{Coverage-Based} $>$ \emph{Test Generation} $>$ \emph{Debugging/Repair}. Ties resolved by this rule are likewise flagged for manual review (\emph{frequency-tie}).
\end{enumerate}

The hierarchy prioritizes categories with stronger methodological specificity and is not intended to imply relative importance among paradigms. Because debugging, repair, and fault-localization papers frequently employ oracle- or assertion-oriented mechanisms within the debugging workflow, frequency-based matching alone may overweight oracle terminology; the semantic override in rule (iii) and the mandatory review of such cases mitigate this risk. Papers with neither title-level matches nor abstract-level frequencies above the threshold are assigned to a residual \emph{Other} category.

For each assignment, the framework computes a confidence level (\emph{High}, \emph{Medium}, or \emph{Low}) based on match frequency and evidence location, with an additional bonus for title-level matches.

\subsection{Noise-Awareness Operationalization}
\label{subsec:noise_operationalization}

The noise-awareness dimension is operationalized to capture not only whether a paper mentions noise-related terminology, but also the \emph{character} of that engagement: explicit modeling of noise phenomena, indirect signaling through hardware-platform mentions, or contradiction by explicit idealization statements. The framework accordingly distinguishes four classes of indicators.

\emph{Strong noise indicators} (e.g., \emph{decoherence}, \emph{readout error}, \emph{depolarizing noise}, \emph{T1/T2 times}, \emph{error mitigation}, \emph{gate errors}) reflect direct engagement with the physical mechanisms of quantum noise and are treated as sufficient evidence of noise-awareness.

\emph{Weak noise indicators} (e.g., \emph{fidelity}, \emph{calibration}, \emph{near-term}, \emph{variational quantum}) are noise-related but appear in both NISQ-aware and idealized contexts. They are treated as evidence of noise-awareness when present, and trigger a noise-review flag when a single weak indicator is the only noise evidence in the paper.

\emph{Hardware-context indicators} (e.g., \emph{IBM Quantum}, \emph{Rigetti}, \emph{IonQ}, \emph{real quantum device}, \emph{quantum hardware}) signal that the work is situated within a practical NISQ context but do not, on their own, constitute direct engagement with noise. They are treated as \emph{indirect} evidence of noise-awareness, sufficient only when not contradicted by explicit idealization markers.

\emph{Idealization indicators} (\emph{noiseless}, \emph{noiseless simulation}, \emph{noise-free}, \emph{ideal simulation}, \emph{ideal simulator}, \emph{idealized execution}, \emph{perfect simulation}, \emph{ideal quantum}, \emph{fault-tolerant}) signal explicit suppression of noise from the work's testing context. When such markers co-occur with hardware-context indicators but no strong noise indicators, the framework classifies the paper as noise-oblivious under the \emph{idealization-override rule}, on the basis that hardware mentions in such papers function as motivational backdrop rather than as evidence of hardware-aware testing.

\begin{figure*}[htbp]
\centering
\resizebox{0.92\textwidth}{!}{%
\begin{tikzpicture}[
  every node/.style={font=\footnotesize},
  decision/.style={diamond, draw, thick, align=center, aspect=2.2,
    inner sep=1pt, fill=blue!8, minimum width=2.9cm, minimum height=1.2cm},
  aware/.style={rectangle, draw, thick, rounded corners, align=center,
    minimum width=1.8cm, minimum height=0.6cm, fill=green!15, font=\footnotesize\bfseries},
  oblivious/.style={rectangle, draw, thick, rounded corners, align=center,
    minimum width=1.8cm, minimum height=0.6cm, fill=red!10, font=\footnotesize\bfseries},
  arrow/.style={-{Latex[length=1.8mm]}, thick},
  lab/.style={font=\scriptsize\itshape}
]
\node[decision] (d1) {Noise keywords\\present?};
\node[decision, right=0.9cm of d1] (d2) {Non-hardware\\indicators?};
\node[decision, right=0.9cm of d2] (d3) {Hardware-only\\indicators?};
\node[decision, right=0.9cm of d3] (d4) {Idealization\\markers?};
\node[aware, right=0.9cm of d4] (out5) {Noise-\\Aware};
\node[oblivious, below=0.7cm of d1] (out1) {Noise-\\Oblivious};
\node[aware,     below=0.7cm of d2] (out2) {Noise-\\Aware};
\node[oblivious, below=0.7cm of d3] (out3) {Noise-\\Oblivious};
\node[oblivious, below=0.7cm of d4] (out4) {Noise-\\Oblivious};
\draw[arrow] (d1) -- node[lab,above]{Yes} (d2);
\draw[arrow] (d2) -- node[lab,above]{No}  (d3);
\draw[arrow] (d3) -- node[lab,above]{Yes} (d4);
\draw[arrow] (d4) -- node[lab,above]{No}  (out5);
\draw[arrow] (d1) -- node[lab,right]{No}  (out1);
\draw[arrow] (d2) -- node[lab,right]{Yes} (out2);
\draw[arrow] (d3) -- node[lab,right]{No}  (out3);
\draw[arrow] (d4) -- node[lab,right]{Yes} (out4);
\end{tikzpicture}}

\caption{Decision process for noise-awareness classification. Any non-hardware noise indicator (strong, other, or weak in Table~\ref{tab:noise_keywords}) is sufficient on its own; hardware-only indicators count as noise-aware unless suppressed by an explicit idealization statement. A single weak indicator as the only evidence, and hardware-only matches, additionally trigger a manual review flag (Section~\ref{subsec:reliability_validation}).}
\label{fig:noise_decision_tree}
\end{figure*}
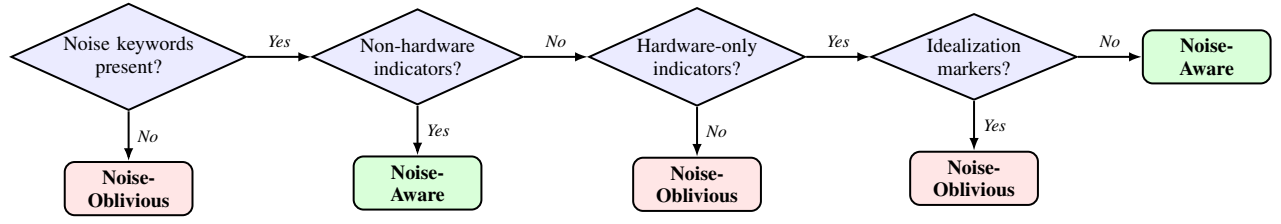

Papers without any valid noise-related evidence are classified as \emph{noise-oblivious}, reflecting an implicit assumption of idealized quantum execution. The decision logic that combines these four classes of indicators is summarized in Figure~\ref{fig:noise_decision_tree}, and the complete keyword framework is summarized in Table~\ref{tab:noise_keywords}.

Three flags drive a noise-review queue: \emph{weak-evidence} (noise-aware on a single weak indicator), \emph{hardware-only-evidence} (noise-aware on hardware mentions alone), and \emph{noise/idealization-conflict} (hardware mentions overridden by idealization markers). Flags do not alter the pipeline output; they direct human attention to the classifications that most warrant inspection.

To capture varying degrees of hardware realism within the noise-aware subset, we further stratify noise-aware papers into three levels: \emph{Strong} (two or more strong indicators); \emph{Moderate} (at least one strong indicator or three or more total indicators); and \emph{Weak} (one or two indicators, of which none are strong). This stratification enables the depth of noise-awareness to be reported alongside its mere presence and supports the discussion of how deeply noise is integrated into noise-aware methodologies (Section~\ref{sec:empirical-landscape}).

\begin{table*}[!t]
\caption{Keyword framework used for classifying noise-awareness. Papers explicitly referencing noise phenomena, NISQ execution, hardware error models, or real quantum hardware platforms are classified as noise-aware. Hardware platform identifiers serve as indirect indicators of practical NISQ contexts unless overridden by explicit idealization statements. Strong, other, and weak terms are each sufficient for a noise-aware label; strength levels (Section~\ref{subsec:noise_operationalization}) depend on the number of strong terms and the total number of matched terms. A single weak term as the only evidence triggers a manual review flag.}
\label{tab:noise_keywords}
\centering
\footnotesize
\begin{tabular}{L{4.2cm} L{10.5cm}}
\toprule
\textbf{Classification} & \textbf{Representative Keywords} \\
\midrule
Noise-Aware (Strong) & noise, noisy, NISQ, decoherence, error mitigation, gate error, measurement error, depolarizing, thermal noise, shot noise, crosstalk, readout error, amplitude damping, phase damping \\
Noise-Aware (Other) & noisy intermediate-scale, hardware error, error rate, coherence time, T1, T2, error model \\
Noise-Aware (Hardware-only, flagged) & IBM Quantum, IBMQ, Rigetti, IonQ, real quantum device, actual hardware, real device, quantum hardware, real quantum hardware, real quantum computer, physical quantum \\
Noise-Aware (Weak, flagged) & fidelity, calibration, near-term, variational quantum \\
Noise-Oblivious & Absence of explicit noise-related terminology, or presence of overriding idealization markers that suppress hardware-only mentions \\
\bottomrule
\end{tabular}
\end{table*}

\subsection{Evaluation-Realism Extraction}
\label{subsec:evaluation_extraction}

To complement the lexical classification described above, the analyses reported in Sections~\ref{subsec:evaluation_realism} and \ref{subsec:practice_patterns} draw on a second, methodologically independent dimension of evidence: the \emph{evaluation configuration} that each contribution explicitly reports. This dimension captures how QST contributions are actually validated, in contrast to the lexical dimension, which captures how they are framed.

The five evaluation-realism attributes used in this paper are the following:

\begin{itemize}
  \item \emph{Backend type}: whether the reported evaluation uses an ideal simulator, a noisy simulator, physical quantum hardware, or any combination thereof; or whether the backend configuration is not explicitly reported.

  \item \emph{Oracle type}: the form of correctness oracle used for test validation (e.g., output-probability oracle with chi-squared test, wrong-output oracle, property-based oracle, dominant-output oracle, or not explicitly reported).

  \item \emph{Statistical evaluation}: whether the paper explicitly reports inferential statistical tests of its experimental results, and if so, which tests (e.g., Mann--Whitney $U$, Vargha--Delaney $\hat{A}_{12}$, Fisher's exact, Wilcoxon signed-rank).

  \item \emph{Artifact availability}: whether the paper provides an explicit replication-artifact link or source-code link.

  \item \emph{Shot budget}: whether the paper explicitly reports the number of measurement shots used (e.g., fixed value, adaptive scheme, or range), and the strategy by which it was determined.
\end{itemize}

\emph{Data source and reporting protocol.}\ For the 59 initial studies, attribute values were taken from the publicly released extraction accompanying~\cite{li2026methodological}, whose extraction schema also defines the attribute categories (e.g., backend and oracle types); for the two added studies, we extracted the same attributes manually using the same schema. In both cases an attribute is recorded as present only when explicitly stated in the paper; ambiguous or underspecified details are recorded as ``not explicitly reported'' rather than inferred. The cross-tabulation of these attributes against the noise-awareness dimension (Sections~\ref{subsec:evaluation_realism} and~\ref{subsec:practice_patterns}) is introduced in this paper.

\subsection{Classification Workflow and Algorithm}
\label{subsec:workflow_and_algorithm}

\begin{figure}[htbp]
\centering
\resizebox{\columnwidth}{!}{
\begin{tikzpicture}[
node distance=0.4cm,
every node/.style={font=\scriptsize},
input/.style={
rectangle, draw=black!70, very thick, rounded corners,
fill=blue!8, minimum width=2.4cm, minimum height=1.8cm,
align=center, inner sep=5pt
},
lexbox/.style={
rectangle, draw=green!50!black, very thick, rounded corners,
fill=green!8, minimum width=3.0cm, minimum height=2.4cm,
align=center, inner sep=5pt
},
evalbox/.style={
rectangle, draw=red!60!black, very thick, rounded corners,
fill=red!8, minimum width=3.0cm, minimum height=2.4cm,
align=center, inner sep=5pt
},
mergebox/.style={
rectangle, draw=purple!60!black, very thick, rounded corners,
fill=purple!8, minimum width=2.8cm, minimum height=1.9cm,
align=center, inner sep=5pt
},
output/.style={
rectangle, draw=black!70, very thick, rounded corners,
fill=cyan!10, minimum width=2.6cm, minimum height=1.9cm,
align=center, inner sep=5pt
},
stagelabel/.style={font=\scriptsize\bfseries, text=black!55},
arrow/.style={-{Latex[length=2.2mm]}, very thick, draw=black!65},
brancharrow/.style={-{Latex[length=2.2mm]}, very thick, draw=black!65}
]
\node[input] (corpus) {
{\Large \faDatabase}\\[4pt]
\textbf{(1) Corpus}\\[3pt]
{\tiny 61 papers}\\
{\tiny (2018--2025)}\\[4pt]
{\tiny \faFile\ Title + Abstract}\\
{\tiny \faKey\ Keywords}
};
\node[lexbox, above right=0cm and 1cm of corpus] (lex) {
{\Large \faSearch}\\[4pt]
\textbf{(2a) Lexical Classification}\\[6pt]
{\tiny \faTags\ Technique assignment}\\
{\tiny \faFilter\ Noise-awareness}\\
{\tiny \faLayerGroup\ Strength stratification}\\
{\tiny \faCheckCircle\ Confidence scoring}\\
{\tiny \faFlag\ Review flags}
};
\node[evalbox, below right=0cm and 1cm of corpus] (eval) {
{\Large \faClipboardCheck}\\[4pt]
\textbf{(2b) Evaluation-Realism}\\[6pt]
{\tiny \faMicrochip\ Backend type}\\
{\tiny \faQuestionCircle\ Oracle type}\\
{\tiny \faChartLine\ Statistical tests}\\
{\tiny \faLink\ Artifact links}\\
{\tiny \faBullseye\ Shot budget}
};
\node[mergebox, right=5.2cm of corpus] (merge) {
{\Large \faProjectDiagram}\\[4pt]
\textbf{(3) Manual Review \&}\\
\textbf{Recording}\\[5pt]
{\tiny \faFlag\ Ambiguity flags}\\
{\tiny \faUserCheck\ Structured review}\\
{\tiny \faFile\ Per-paper log}
};
\node[output, right=1cm of merge] (out) {
{\Large \faChartBar}\\[4pt]
\textbf{(4) Findings}\\[5pt]
{\tiny \faSitemap\ Taxonomy}\\
{\tiny \faChartBar\ Trends}\\
{\tiny \faCheckDouble\ Evaluation realism}\\
{\tiny \faExclamationTriangle\ Gaps}
};
\node[stagelabel] at (corpus.north) [yshift=.75cm] {INPUT};
\node[stagelabel, above=0.4cm of merge.north] {PROCESSING};
\node[stagelabel, above=0.4cm of out.north]   {OUTPUT};

\draw[brancharrow] (corpus.east) -- ++(0.45,0) |- (lex.west);
\draw[brancharrow] (corpus.east) -- ++(0.45,0) |- (eval.west);
\draw[brancharrow] (lex.east) -- ++(0.45,0) |- (merge.west);
\draw[brancharrow] (eval.east) -- ++(0.45,0) |- (merge.west);
\draw[arrow] (merge.east) -- (out.west);

\node[font=\tiny\itshape, above=0.08cm of lex.north,
text=green!40!black]
{Lexical (text-based) line of evidence};
\node[font=\tiny\itshape, below=0.25cm of eval.south, text=red!50!black]
  {Evaluation-realism (metadata-based) line of evidence};

\end{tikzpicture}
}
\caption{Hierarchical workflow of the two-pipeline methodology. The lexical classification pipeline applies regex-based matching for testing-technique and noise-awareness assignment, and raises review flags for ambiguous cases. The evaluation-realism attributes (backend type, oracle type, statistical-test usage, shot budget, and artifact availability) are sourced according to the protocol described in Section~\ref{subsec:evaluation_extraction}, which records each attribute only when explicitly stated. Both streams converge at the manual-review-and-recording stage, where flagged papers are inspected, final categories are recorded, and supporting rationale is logged.}
\label{fig:classification_workflow}
\end{figure}
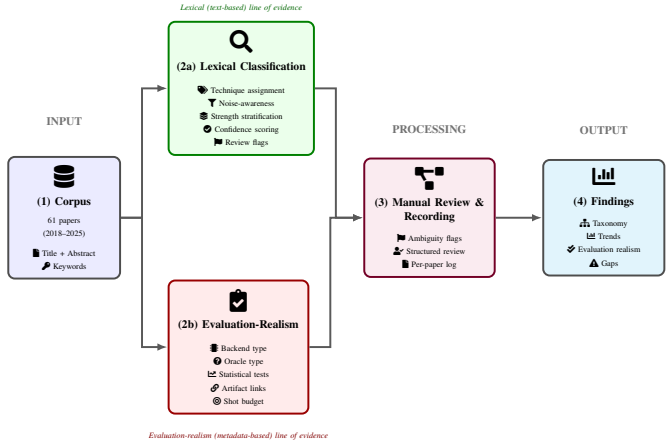

The classification pipeline (Figure~\ref{fig:classification_workflow}) proceeds through three phases, formalized in Algorithm~\ref{alg:classification}. \emph{Phase~I} assigns the provisional testing category using the rules of Section~\ref{subsubsec:tie_breaking} and emits the matched keywords, confidence level, and any testing flags. \emph{Phase~II} applies the four-class noise-indicator logic of Section~\ref{subsec:noise_operationalization}, assigns the strength level, and emits any noise flags. \emph{Phase~III} subjects all flagged papers to the structured review of Section~\ref{subsec:reliability_validation}; any override is recorded with its written rationale and replayed identically on every subsequent run. The pipeline is therefore fully deterministic: the same keyword sets, thresholds, decision rules, and review log applied to the same metadata yield identical final classifications.

\begin{algorithm*}[htbp]
\caption{Deterministic Taxonomy-Guided Classification with
Manual Review Integration}
\label{alg:classification}
\footnotesize\linespread{0.94}\selectfont
\KwIn{Dataset $D$ containing paper metadata; threshold $\tau$;
manual review decisions $R$ (initially empty)}
\KwOut{Testing category, noise-awareness label, confidence score,
strength level, review flags, and final category after manual
review}
Initialize testing lexicons $\mathcal{K}_T$\;
Initialize noise lexicons: strong ($\mathcal{K}_S$), hardware
($\mathcal{K}_H$), weak ($\mathcal{K}_W$), idealization
($\mathcal{K}_I$)\;
Initialize ambiguity terms $\mathcal{K}_A$\;
Precompile case-insensitive word-boundary regex patterns\;
\BlankLine
\ForEach{$p \in D$}{
Construct unified corpus $\mathcal{C}_p$ from title, abstract,
and Bib\TeX{} keywords (lowercased)\;
\BlankLine
\textbf{Phase~I: Testing Classification}\;
Extract title-level matches per category\;
\If{title matches more than one category}{
  Flag $p$ for manual review (multiple-title-match)\;
}
\If{title contains both debugging and oracle/assertion keywords}{
  Provisionally assign $p$ to \emph{Debugging/Repair}
  (semantic override)\;
  Flag $p$ for manual review (oracle+debugging-overlap)\;
  Skip remainder of Phase~I\;
}
\eIf{title matches exist}{
  Assign category by priority hierarchy among title matches\;
  Apply title-weight confidence bonus\;
}{
  Detect ambiguity terms $\mathcal{K}_A$ in $\mathcal{C}_p$;
  flag if present\;
  Compute corpus-level keyword frequencies $F_T$\;
  \eIf{$\max(F_T) \geq \tau$}{
    \If{tie among multiple categories at $\max(F_T)$}{
      Flag $p$ for manual review (frequency-tie)\;
    }
    Assign category by priority hierarchy among tied categories\;
    Compute confidence score\;
  }{
    Assign category: \emph{Other}\;
  }
}
\BlankLine
\textbf{Phase~II: Noise Classification}\;
Detect strong and other ($\mathcal{I}_S$), hardware-only
($\mathcal{I}_H$), weak ($\mathcal{I}_W$), and idealization
($\mathcal{I}_I$) indicators in $\mathcal{C}_p$\;
\uIf{$\mathcal{I}_S \neq \emptyset$ or $\mathcal{I}_W \neq \emptyset$}{
  Assign: \emph{Noise-Aware}\;
  \If{the only matched indicator is a single weak indicator}{
    Flag $p$ for noise review (weak-evidence)\;
  }
}\uElseIf{$\mathcal{I}_H \neq \emptyset$ and
$\mathcal{I}_I = \emptyset$}{
  Assign: \emph{Noise-Aware}\;
  Flag $p$ for noise review (hardware-only-evidence)\;
}\uElseIf{$\mathcal{I}_H \neq \emptyset$ and
$\mathcal{I}_I \neq \emptyset$}{
  Assign: \emph{Noise-Oblivious} (idealization-override)\;
  Flag $p$ for noise review (noise/idealization-conflict)\;
}\Else{
  Assign: \emph{Noise-Oblivious}\;
}
Estimate strength level: Weak / Moderate / Strong\;
\BlankLine
\textbf{Phase~III: Manual Review Integration and Recording}\;
\If{$p$ is flagged in Phase~I or Phase~II, or $p$ is
Noise-Oblivious with $\mathcal{I}_I \neq \emptyset$}{
  Add $p$ to the manual-review queue\;
  Apply structured review\;
  \If{review yields a final primary category different from
  provisional assignment}{
    Override $p$'s testing category with $R[p].\text{final\_category}$\;
    Record reassignment rationale from $R[p]$\;
  }
}
Record: pipeline category, final category, noise label,
confidence, strength level, matched indicators, idealization
evidence, all raised flags, and manual-review decision (where
applicable)\;
}
\BlankLine
\Return classified dataset with ambiguity annotations and
manual-review log\;
\end{algorithm*}

\subsection{Reliability and Validation}
\label{subsec:reliability_validation}

A manual validation layer addresses four classes of ambiguity in lexical classification (category ties, low-confidence assignments, non-canonical terminology, and noise evidence that may be peripheral rather than substantive) while preserving the reproducibility of the pipeline.

\emph{Hierarchical tie resolution.} Ties are resolved by the priority hierarchy of Section~\ref{subsubsec:tie_breaking}, so every tie has a single, auditable outcome.

\emph{Automated ambiguity flagging.} The pipeline flags papers for manual review whenever any of six conditions holds: (i)~the title matches keywords from more than one category (\emph{multiple-title-match}); (ii)~the title triggers the debugging/assertion semantic override (\emph{oracle+debugging-overlap}); (iii)~a tie among categories must be broken by the priority hierarchy (\emph{frequency-tie}); (iv)~the title yields no category match and the unified text contains ambiguous terminology, matched as plain substrings (e.g., \emph{statistical testing}, \emph{black-box testing}, \emph{runtime assertion}, \emph{monitoring}, \emph{verification}, \emph{validation}, \emph{equivalence checking}) that frequently spans multiple paradigms; (v)~the paper's noise classification rests on weak or hardware-only evidence; or (vi)~hardware-only mentions co-occur with idealization markers (the \emph{noise/idealization-conflict} case). In addition, papers classified as noise-oblivious whose text contains idealization markers are added to the testing-category review queue without raising a flag, so that their idealized scope is confirmed.

\emph{Structured manual review.} Each flagged paper undergoes targeted full-text inspection. For every such paper the reviewer records four items: the final primary category; a secondary relation (the testing concept the paper engages with as a supporting mechanism, if any); a written rationale for the assignment decision; and an indication of whether the manual review confirmed or overrode the pipeline's provisional category. The complete review log, with per-paper rationale, is retained and reapplied to all subsequent reproductions of the classification.

\emph{Noise-classification review.} Papers flagged on the noise dimension (weak, hardware-only, or conflict cases) are inspected for two specific patterns: (i)~whether weak or hardware-only lexical evidence corresponds to substantive engagement with NISQ-era hardware noise in the paper body; and (ii)~whether idealization markers (e.g., \emph{fault-tolerant}) genuinely override hardware mentions or instead coexist with substantive NISQ engagement. The manual reviewer either confirms the pipeline classification or records a justified override with written rationale.

\subsubsection{Empirical Reliability of the Classification Framework}
\label{subsubsec:framework_reliability}

Beyond describing the manual-review mechanism, the resulting review log permits a direct empirical assessment of how reliably the deterministic pipeline reproduces the final, manually validated classifications. We report four reliability measures derived from comparing pipeline-assigned categories against post-review final categories across the 61-paper corpus. Table~\ref{tab:framework_reliability} summarizes the results.

\begin{table}[htbp]
\caption{Agreement between pipeline and final (post-review) classifications. Non-flagged papers are those that raised no testing-category flag, including the two papers queued by the idealization-marker rule; flagged papers are the 11 that raised one. Noise stability refers to the 6 papers flagged on the noise dimension.}
\label{tab:framework_reliability}
\centering
\footnotesize
\begin{tabular}{l c c}
\toprule
\textbf{Reliability measure}                & \textbf{Result} & \textbf{Rate} \\
\midrule
Overall pipeline--final agreement (testing) & 56 / 61         & 91.8\% \\
Agreement among non-flagged papers          & 50 / 50         & 100.0\% \\
Agreement among flagged papers              & 6 / 11          & 54.5\% \\
Reassignment rate within flagged set        & 5 / 11          & 45.5\% \\
Reassignment rate across full corpus        & 5 / 61          & 8.2\% \\
\midrule
Noise classification stability              & 6 / 6           & 100.0\% \\
\bottomrule
\end{tabular}
\end{table}

Across the 61-paper dataset, the pipeline assigned 52 papers a High-confidence testing category and one a Medium-confidence category; the remaining 8 fell below the evidence threshold and were assigned to \emph{Other}. Thirteen papers entered the testing-category review queue: eleven were flagged automatically under one or more of the conditions above, and two additional papers~\cite{paper_37, paper_112} were queued by the idealization-marker rule without raising an explicit flag.

Of the thirteen queued papers, eight were confirmed in their pipeline-assigned category and five were reassigned following semantic review. All five reassignments fell within the eleven automatically flagged papers, giving a reassignment rate of 45.5\% within the auto-flagged subset (Table~\ref{tab:framework_reliability}) and 38.5\% across the full review queue of thirteen, or 8.2\% across the full corpus. Six additional papers entered the noise-classification review queue; all six pipeline classifications were retained, with the flags serving to surface the underlying evidence type rather than to trigger reassignment.

Three observations follow from these measures. First, the pipeline's category assignment is highly stable on the unambiguous portion of the corpus: every paper the pipeline classified without raising any ambiguity flag (50 of 61) retained its pipeline-assigned category after manual review, yielding a 100\% confirmation rate on the non-flagged set. This 100\% figure is in part a structural property of the workflow rather than an independent accuracy estimate: apart from the two papers queued by the idealization-marker rule, papers that raised no ambiguity flag were not routed to manual review, so by construction they had no opportunity to be reassigned. It should therefore be read as evidence that the flag conditions capture the cases worth reviewing, not as a direct measure of classifier accuracy on the non-flagged set. Second, the flags are selective rather than indiscriminate: nearly half of the flagged papers (5 of 11, 45.5\%) were reassigned, indicating that the flag conditions isolate cases in which the lexical evidence genuinely warrants human inspection. Third, the flags are not exhaustive: in the external validation (Section~\ref{subsubsec:interrater_validation}), one paper that raised no flag~\cite{paper_94} was adjudicated to a different category, so a small residual error rate remains among non-flagged papers. The noise-classification pipeline exhibits even greater stability, with zero of the six flagged classifications overridden after review. Taken together, these measures indicate that the deterministic pipeline reproduces the final classifications for the large majority of the corpus without manual intervention, while the ambiguity-flagging layer reliably isolates the cases in which manual judgment changes the outcome.

\subsubsection{Inter-Rater Reliability and Ground-Truth Validation}
\label{subsubsec:interrater_validation}

Although the preceding analysis evaluates the stability of the deterministic classification framework, it does not establish whether the taxonomy itself can be applied consistently by independent reviewers. To assess the reproducibility of the proposed taxonomy and to construct an external reference dataset for classifier validation, we conducted an inter-rater agreement study on a validation sample of 31 papers drawn from the analyzed corpus.

The validation sample was selected to provide coverage across the testing-taxonomy categories rather than as a simple proportional draw. It combines a random validation subset with papers from targeted inclusion strata. Because the sample is stratified, the agreement rates reported below characterize reliability on this validation set and should not be interpreted as corpus-level estimates without reweighting to the class distribution of the full dataset. Each paper was independently classified by two reviewers using only the title, abstract, and author-provided keywords. The first reviewer was the author of the original classification labels; the second reviewer had not been involved in building the pipeline. This restriction mirrors the information available to the automated classification pipeline and therefore enables a direct comparison between human and automated classification decisions under identical evidence conditions.

The reviewers assigned labels along the two classification dimensions considered in this study: (i) testing-technique category and (ii) noise-awareness classification. To reduce confirmation bias, the second reviewer was blinded to both the automated classifications and the coding decisions of the first reviewer throughout the validation exercise. Both reviewers applied the same operational definitions, category descriptions, and decision rules used by the deterministic classification framework.

Inter-rater agreement was assessed separately for the testing-category and noise-awareness dimensions using Cohen's kappa. For testing-technique classification, the reviewers achieved almost perfect agreement ($\kappa = 0.962$), indicating that the taxonomy definitions can be applied consistently and reproducibly by independent annotators. For the noise-awareness dimension, the reviewers achieved substantial agreement ($\kappa = 0.760$), providing evidence that the operational criteria introduced in Section~\ref{subsec:noise_operationalization} yield stable classifications when applied by different reviewers. Taken together, these results suggest that both dimensions of the framework exhibit a high degree of conceptual clarity and operational consistency.

Following the independent coding phase, the reviewers disagreed on one testing-category label and three noise labels. These disagreements were resolved through structured adjudication: the reviewers jointly re-examined the disputed studies, and category decisions were recorded together with their supporting rationale in an adjudication log. This process produced a consensus classification for every study in the validation sample and established the adjudicated ground truth used for classifier evaluation.

We evaluated the framework against this adjudicated ground truth in two stages that mirror its design. In the first stage, we compared the \emph{raw automated classifier} (the deterministic rule-based output before any manual review) against the ground truth. For testing-technique classification, the automated classifier agreed with the ground truth for 24 of the 31 validation papers (77.4\%), corresponding to substantial agreement ($\kappa = 0.737$). For the noise-awareness dimension, the automated classifier agreed with the ground truth for 29 of the 31 papers (93.5\%), achieving almost perfect agreement ($\kappa = 0.852$); notably, this exceeded the agreement observed between independent human reviewers ($\kappa = 0.760$), indicating that the operationalized noise-awareness criteria can be reproduced reliably through deterministic rule-based classification.

In the second stage, we applied the framework's manual-review layer (targeted inspection of the cases the pipeline itself flagged as ambiguous) and compared the resulting \emph{final} classification against the ground truth. The manual review corrected five testing-category assignments (papers reassigned from the pipeline's provisional label to the adjudicated category), raising testing-technique agreement to 29 of the 31 papers (93.5\%, $\kappa = 0.922$). The noise dimension required no manual reassignment, so the final noise classification is identical to the automated result ($\kappa = 0.852$). The gap between the two stages on the testing dimension ($\kappa = 0.737$ versus $\kappa = 0.922$) quantifies the accuracy contributed by reserving manual inspection for automatically flagged cases.

The residual disagreements are retained in the corpus-level analysis as produced by the framework, so that the reported results reflect its output rather than post hoc corrections. On the testing dimension, two papers adjudicated as Oracle/Assertion-Based Testing remain under Test Generation~\cite{paper_94} and Debugging/Repair~\cite{li2025statistical}. On the noise dimension, the two residual disagreements run in opposite directions~\cite{paper_36, paper_122}, so applying the adjudicated labels would leave the corpus-wide count of noise-aware papers unchanged at 14.

Taken together, the inter-rater results and the staged classifier-validation results provide complementary evidence of reliability. The inter-rater agreement demonstrates that the taxonomy can be applied consistently by independent human reviewers; the automated classifier reproduces those human judgments at substantial-to-almost-perfect levels without any manual intervention; and the targeted manual-review layer closes most of the residual gap on the testing dimension.

\begin{table}[htbp]
\caption{Summary of inter-rater and staged classifier-validation results. Stage~1 reports the raw deterministic classifier against the adjudicated ground truth; Stage~2 reports the framework's final classification after manual review of automatically flagged cases.}
\label{tab:interrater_validation}
\centering
\footnotesize
\resizebox{\columnwidth}{!}{%
\begin{tabular}{lcc}
\toprule
\textbf{Validation Measure}                        & \textbf{Testing} & \textbf{Noise} \\
\midrule
Human--Human Cohen's $\kappa$                      & 0.962            & 0.760 \\
\midrule
Stage 1: Automatic classifier--GT agreement        & 24/31 (77.4\%)   & 29/31 (93.5\%) \\
Stage 1: Automatic classifier--GT Cohen's $\kappa$ & 0.737            & 0.852 \\
\midrule
Stage 2: Final (post-review)--GT agreement         & 29/31 (93.5\%)   & 29/31 (93.5\%) \\
Stage 2: Final (post-review)--GT Cohen's $\kappa$  & 0.922            & 0.852 \\
\bottomrule
\end{tabular}}
\end{table}

This hybrid design extends the adaptive-reading-depth guideline articulated for systematic mapping studies in software engineering~\cite{petersen2008systematic}, in which classification effort is concentrated on papers whose categorization is not immediately clear, by replacing the initial abstract-reading pass with a deterministic lexical pipeline and reserving manual inspection for cases the pipeline itself flags as ambiguous.

\section{Empirical Findings}
\label{sec:empirical-landscape}

This section reports the findings for the four research questions of Section~\ref{subsec:research_design}; each subsection closes with a summary answer. Throughout, the label \emph{noise-aware} denotes explicit terminological engagement with noise as captured by the lexical framework of Section~\ref{subsec:noise_operationalization}, which reflects how a contribution is \emph{framed}. Whether a contribution is \emph{validated} under hardware-realistic conditions is measured separately through its reported execution backends (Section~\ref{subsec:evaluation_realism}). The findings derive their force from the convergence of the two dimensions rather than from either in isolation.

\subsection{RQ1: Prevalence of Noise-Awareness}
\label{subsec:rq1}

\subsubsection{Lexical Noise-Awareness}
\label{subsec:overall_distribution}

\begin{figure}[htbp]
\centering
\includegraphics[width=0.72\columnwidth]{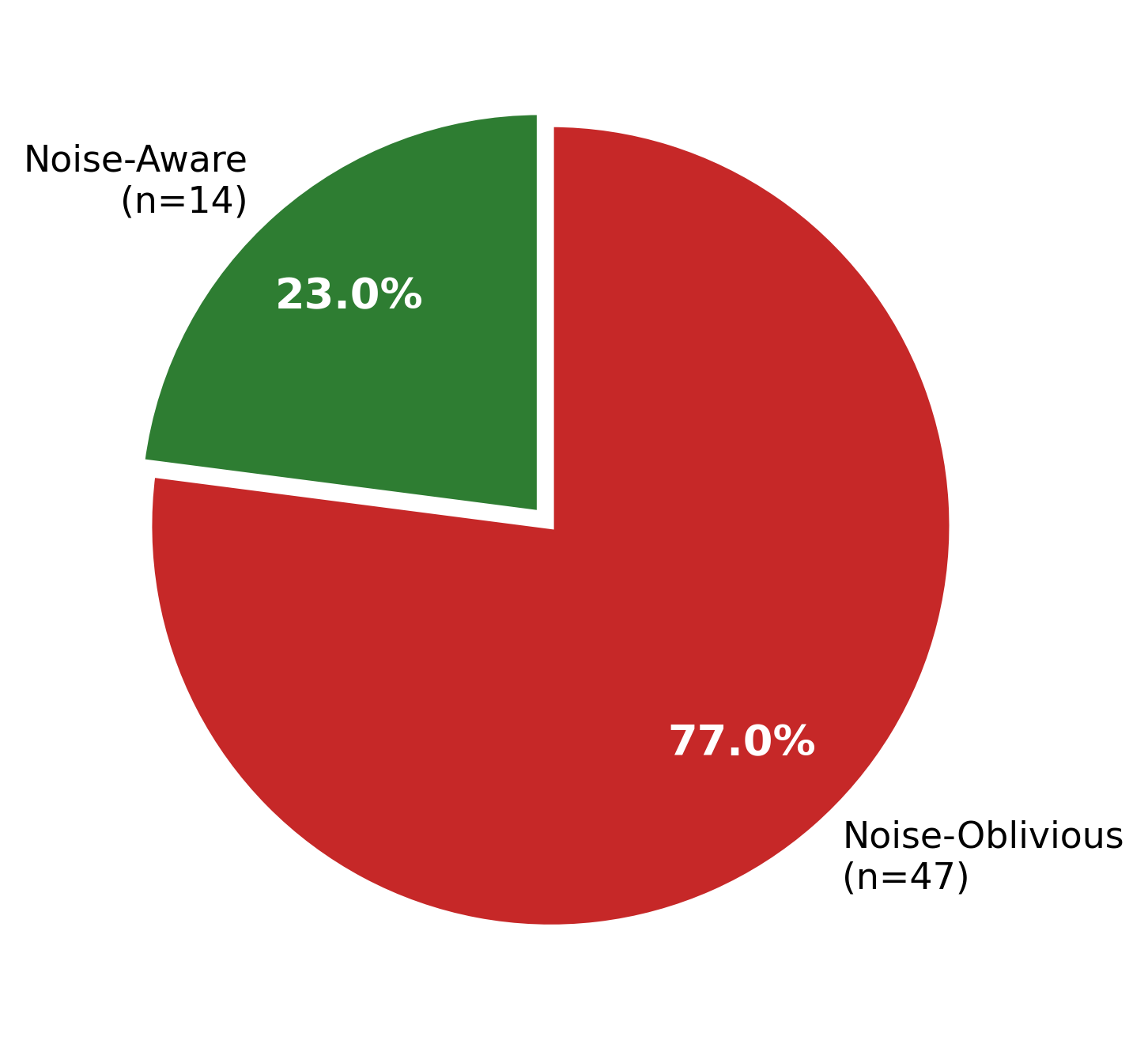}
\caption{Overall distribution of noise-awareness in quantum software testing research ($N=61$, 2018--2025). Of the analyzed papers, 47 (77.0\%) are noise-oblivious and 14 (23.0\%) explicitly consider noise or NISQ hardware constraints.}
\label{fig:overall_distribution}
\end{figure}

Only 14 of the 61 analyzed papers (23.0\%) explicitly address noise or NISQ execution environments, while the remaining 47 (77.0\%) adopt noise-oblivious assumptions (Wilson 95\% confidence interval for the noise-aware share: [14.2\%, 34.9\%]). The resulting ratio of 3.36:1 indicates that approximately three out of every four QST contributions continue to operate under idealized models of quantum computation, largely abstracting away the practical constraints imposed by contemporary NISQ hardware~\cite{preskill2018nisq, yousuf2026systematic} (Figure~\ref{fig:overall_distribution}).

The framework's stratification of noise-aware papers into Strong, Moderate, and Weak levels reveals additional nuance. Among the 14 noise-aware papers, 5 (35.7\% of the noise-aware subset) provide strong noise integration with multiple direct noise indicators, for example, explicit references to decoherence, error mitigation, or measurement errors. A further 4 papers (28.6\%) provide moderate integration, and the remaining 5 (35.7\%) provide only weak integration based on a small number of hardware-platform or peripheral noise mentions.

Expressed against the full dataset, only 8.2\% of papers exhibit strong noise integration, indicating that deep methodological engagement with noise remains rare even within the noise-aware subset.

\subsubsection{Evaluation Realism from Reported Backends}
\label{subsec:evaluation_realism}

Table~\ref{tab:backend_distribution} summarizes the distribution of reported evaluation backends across the 61-paper corpus. Ideal-simulator-only evaluation dominates, accounting for 43 papers (70.5\%). A further 7 papers (11.5\%) do not explicitly report their evaluation backend, leaving the configuration underspecified. Aggregated, only 11 of 61 papers (18.0\%) explicitly report any evaluation under hardware-realistic conditions (physical hardware or a noisy simulator; Wilson 95\% confidence interval [10.4\%, 29.5\%]), and only 5 papers (8.2\%) report evaluation on physical quantum hardware.

\begin{table}[htbp]
\caption{Reported execution backends of the analyzed papers. Categories follow the extraction schema of~\cite{li2026methodological}; papers that do not state their backend are listed as \emph{not explicitly reported}.}
\label{tab:backend_distribution}
\centering
\footnotesize
\resizebox{\columnwidth}{!}{%
\begin{tabular}{l r r}
\toprule
\textbf{Reported Backend Configuration}                         & \textbf{N}  & \textbf{\%} \\
\midrule
Ideal simulator only                                            & 43          & 70.5 \\
Ideal simulator + noisy simulator                               & 5           & 8.2 \\
Ideal simulator + physical hardware                             & 2           & 3.3 \\
Physical hardware + noisy simulator                             & 2           & 3.3 \\
Noisy simulator only                                            & 1           & 1.6 \\
Physical hardware only                                          & 1           & 1.6 \\
Not explicitly reported                                         & 7           & 11.5 \\
\midrule
\textbf{Total}                                                  & \textbf{61} & \textbf{100.0} \\
\midrule
\multicolumn{3}{l}{\textbf{Derived aggregates:}} \\
\quad Hardware-realistic (physical hardware or noisy simulator) & 11          & 18.0 \\
\quad Physical hardware (any combination)                       & 5           & 8.2 \\
\quad Ideal-simulator-based (any inclusion of ideal simulator)  & 50          & 82.0 \\
\bottomrule
\end{tabular}}
\end{table}

This direct evaluation-realism statistic complements the lexical noise-awareness finding (23.0\% noise-aware by terminology) and indicates that even the lexical rate may slightly overstate the proportion of QST contributions whose validation is grounded in hardware-realistic execution.

The two numbers reported so far, 47 noise-oblivious papers (lexical) and 43 ideal-simulator-only papers (evaluation realism), are close in magnitude but measure different properties of the corpus and therefore do not coincide. Table~\ref{tab:noise_vs_backend} reports their joint distribution and clarifies the relationship. Of the 47 noise-oblivious papers, 38 evaluate exclusively on an ideal simulator and 7 do not explicitly report an evaluation backend. Two run on hardware-realistic backends despite the absence of noise-related terminology in their abstracts and keywords: the dynamic runtime-assertion framework~\cite{paper_122} combines an ideal simulator with physical hardware, and the statistical bug-locating method~\cite{sato2025bug} combines an ideal simulator with a noisy simulator. Conversely, 5 of the 14 noise-aware papers evaluate exclusively on an ideal simulator: the entanglement-guided neural-network testing framework \emph{QuanTest}~\cite{paper_36}; the hybrid-distance random testing strategy~\cite{paper_52}; the multi-subroutine unit-to-integration testing work~\cite{paper_71}; \emph{QCRMut}~\cite{paper_113}; and the statistical bug/noise-discrimination approach~\cite{paper_116}. These five engage with noise terminologically without grounding their validation in hardware-realistic execution. The 43-paper figure in Table~\ref{tab:backend_distribution} is therefore not the noise-oblivious set restricted to ideal-simulator evaluation; it is the larger union of two distinct groups: 38 noise-oblivious papers whose lexical framing and evaluation backend agree, and 5 noise-aware papers whose lexical framing and evaluation backend disagree.

\begin{table}[htbp]
\caption{Noise-awareness classification versus reported evaluation backend. Hardware-realistic denotes any configuration that includes physical hardware or a noisy simulator.}
\label{tab:noise_vs_backend}
\centering
\footnotesize
\begin{tabular}{l c c c c}
\toprule
& \textbf{Ideal}     & \textbf{Hardware-} & \textbf{Not}      & \textbf{Total} \\
& \textbf{sim. only} & \textbf{realistic} & \textbf{reported} &  \\
\midrule
Noise-Aware     & 5                  & 9                  & 0                 & 14 \\
Noise-Oblivious & 38                 & 2                  & 7                 & 47 \\
\midrule
\textbf{Total}  & \textbf{43}        & \textbf{11}        & \textbf{7}        & \textbf{61} \\
\bottomrule
\end{tabular}
\end{table}

This decomposition is itself a substantive finding: 5 of the 14 lexically noise-aware papers (35.7\% of the noise-aware subset) do not back up their noise-aware framing with a hardware-realistic evaluation, while 2 of the 47 lexically noise-oblivious papers quietly run on hardware-realistic backends. The two dimensions of analysis are therefore not redundant: each surfaces methodological misalignments that the other obscures.

\paragraph*{Convergence of the two dimensions} Despite these individual mismatches, the two dimensions are strongly associated at the corpus level. Hardware-realistic evaluation is reported by 9 of the 14 lexically noise-aware papers (64.3\%) but by only 2 of the 47 noise-oblivious papers (4.3\%) (Table~\ref{tab:noise_vs_backend}; Fisher's exact test, odds ratio 40.5, $p < 0.001$). Equivalently, 9 of the 11 papers with hardware-realistic evaluation are classified as noise-aware, and 45 of the 50 papers without it are classified as noise-oblivious. The share also rises with the strength of noise integration: 2 of 47 noise-oblivious papers, 2 of 5 Weak, 3 of 4 Moderate, and 4 of 5 Strong papers report hardware-realistic evaluation (Cochran--Armitage test for trend across these four levels, $z = 5.33$, $p < 0.001$), although within the noise-aware subset the difference between Strong or Moderate and Weak papers is not statistically significant (Fisher's exact $p \approx 0.27$). Because the evaluation attributes were extracted from the full papers independently of the abstract-level classifier, this association provides convergent evidence that the lexical noise-awareness label tracks how contributions are actually evaluated.

Five papers in the dataset explicitly report evaluation on physical quantum hardware. The earliest, on dynamic runtime assertions~\cite{paper_122}, appeared in 2020 and combined ideal-simulator and hardware execution; an online-monitoring study using test points on three IBM~Q devices (Melbourne, Cambridge, and Paris) followed in 2021~\cite{paper_15} and additionally reports noisy-simulator evaluation. A characterization of quantum-noise impact on quantum programs appeared in 2023~\cite{paper_67} and combines physical hardware with noisy-simulator evaluation, while a 2024 study tests quantum programs through commuting Pauli strings on IBM hardware~\cite{paper_39} and is the only contribution to report physical hardware exclusively. The most recent, \emph{HornBro}~\cite{paper_44} (2025), combines ideal-simulator and physical-hardware evaluation. The remainder of the hardware-realistic subset relies on noisy simulators rather than on real devices.

\begin{answerbox}{Answer to RQ1}
Noise-awareness is a minority concern in QST research. Only 23.0\% of papers (14 of 61) engage with noise at the framing level, only 8.2\% exhibit strong noise integration, and only 18.0\% report any hardware-realistic evaluation. Framing and validation diverge for individual papers (5 of the 14 noise-aware papers evaluate exclusively on ideal simulators, while 2 noise-oblivious papers run on hardware-realistic backends), yet are strongly associated overall (odds ratio 40.5, $p < 0.001$).
\end{answerbox}

\subsection{RQ2: Variation Across Testing Paradigms}
\label{subsec:by_technique}

\begin{figure}[htbp]
\centering
\includegraphics[width=\columnwidth]{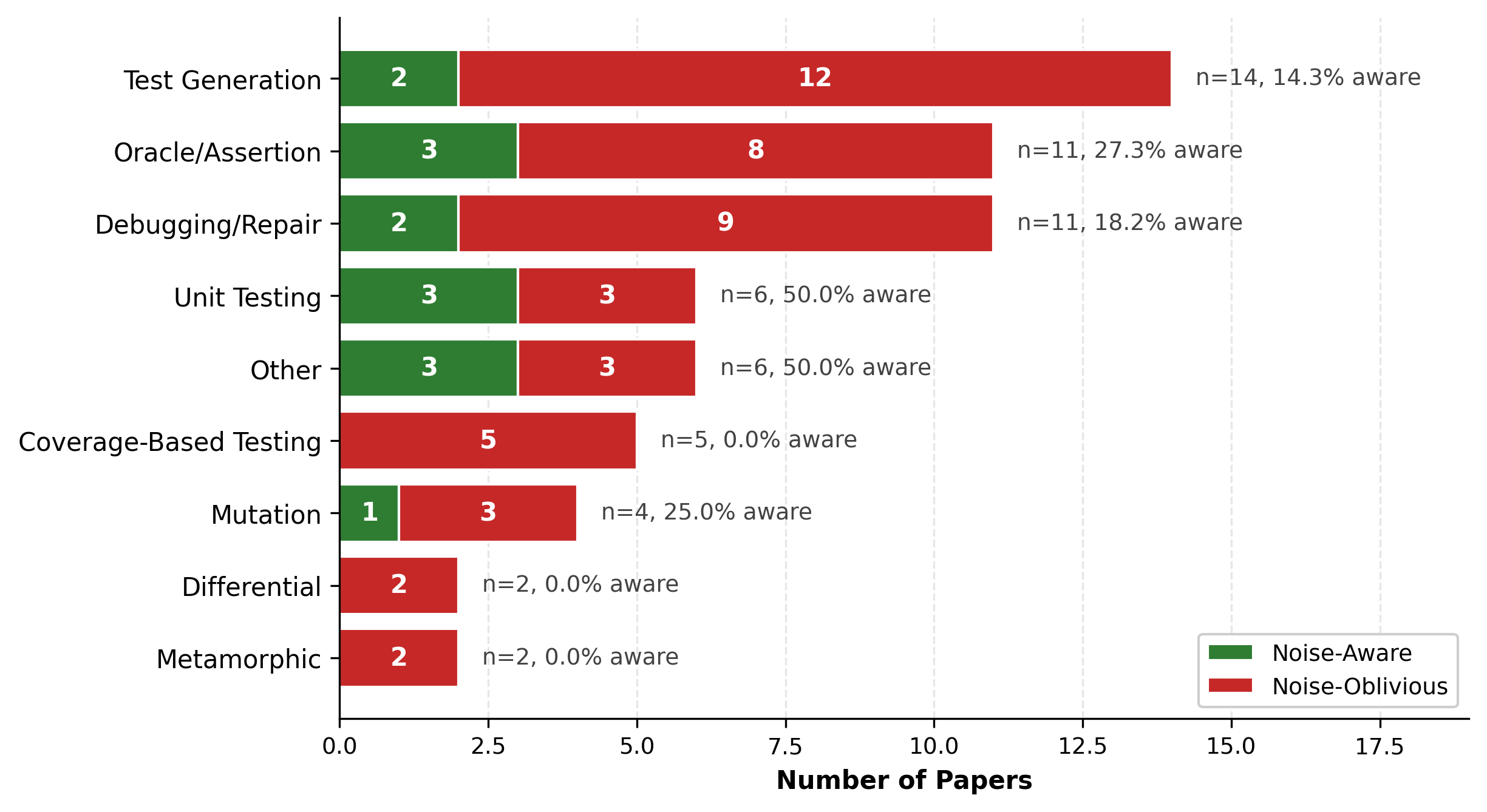}
\caption{Distribution of noise-aware (green) and noise-oblivious (red) studies across eight primary testing paradigms and a residual \emph{Other} category. Test generation dominates with 14 papers, of which only 2 explicitly consider noise. Unit testing and the residual \emph{Other} category exhibit the highest noise-awareness rates (3 of 6 papers each, 50.0\%). Coverage-based testing (0/5), metamorphic testing (0/2), and differential testing (0/2) contain no noise-aware contributions in the analyzed corpus.}
\label{fig:technique_breakdown}
\end{figure}

Figure~\ref{fig:technique_breakdown} presents the distribution of noise-aware and noise-oblivious studies across the eight primary testing paradigms and the residual \emph{Other} category.

Test generation remains the most extensively studied paradigm, with 14 papers (23.0\% of the dataset), yet only 2 (14.3\%) explicitly consider noise. Thirteen of the fourteen evaluate on ideal simulators only. Oracle and assertion-based testing (11 papers, 27.3\% noise-aware) and debugging, repair, and fault localization (11 papers, 18.2\% noise-aware) are equally well-represented as the next most studied paradigms, but their noise-awareness profiles differ: oracle/assertion-based testing exhibits the third-highest noise-awareness rate among major paradigms, while debugging/repair remains comparatively idealized.

Mutation testing remains predominantly noise-oblivious, with 3 of 4 studies (75.0\%) classified as noise-oblivious, and coverage-based testing is entirely noise-oblivious (0 of 5 papers).

A notable pattern is the emergence of unit testing of quantum subroutines and the residual \emph{Other} category as the most hardware-aware paradigms, each at 50.0\% noise-awareness (3 of 6 papers in each case; small-sample caveat applies; see Section~\ref{subsec:disc_gap} for an interpretation). Metamorphic testing (2 papers, 0\% noise-aware) and differential testing (2 papers, 0\% noise-aware) remain underexplored, with no contributions explicitly addressing noisy execution conditions in our dataset.

Because several paradigms rest on small subsamples, we report Wilson 95\% confidence intervals alongside the point estimates (Table~\ref{tab:comparative_synthesis}) and test the paradigm$\times$noise-awareness association inferentially. The apparent paradigm ranking is not statistically distinguishable at this sample size: the unit-testing rate of 50.0\% (3/6) carries a 95\% Wilson interval of $[19\%, 81\%]$, which overlaps almost entirely with the $[4\%, 40\%]$ interval of the 14.3\% (2/14) test-generation rate. A Freeman--Halton extension of Fisher's exact test over the full paradigm$\times$noise-awareness contingency table yields $p \approx 0.40$, and a pairwise Fisher's exact test between unit testing and test generation yields $p \approx 0.13$; neither rejects the null of no association. The per-paradigm rates should therefore be read as descriptive, and the paradigm-level contrasts as suggestive rather than statistically established.

\subsubsection{Volume versus Noise-Awareness}
\label{subsec:synthesis}

\begin{figure}[htbp]
\centering
\includegraphics[width=\columnwidth]{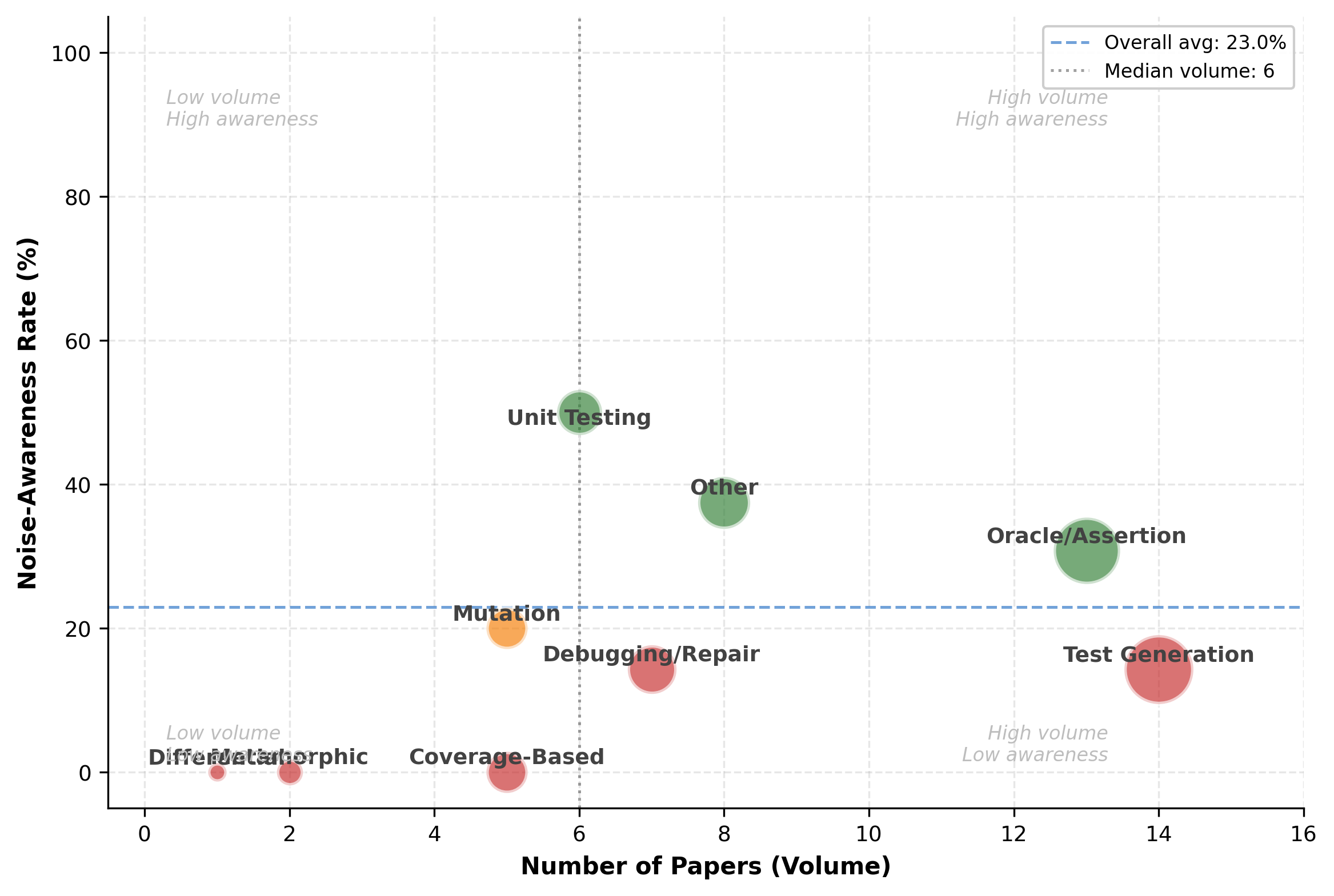}
\caption{Testing paradigms positioned by publication volume (x-axis) and noise-awareness rate (y-axis). Bubble size reflects paper count. Dashed reference lines partition the plane into four quadrants at the corpus median volume ($n = 6$ papers) and the corpus-wide noise-awareness rate (23.0\%). The lower-right quadrant (volume above median, awareness below corpus rate) contains Test Generation and Debugging/Repair. The upper-right quadrant is occupied solely by Oracle/Assertion-Based Testing (27.3\%). The upper-left quadrant contains Unit Testing (50.0\%), the residual \emph{Other} category (50.0\%), and Mutation Testing (25.0\%). The lower-left quadrant contains Coverage-Based, Metamorphic, and Differential Testing, all at zero noise-awareness.}
\label{fig:volume_vs_awareness}
\end{figure}

To synthesize the paradigm-level findings, Figure~\ref{fig:volume_vs_awareness} positions each testing paradigm along two dimensions: publication volume (x-axis) and noise-awareness rate (y-axis). Dashed reference lines partition the plane at the corpus median volume ($n = 6$ papers) and the corpus-wide noise-awareness rate (23.0\%), revealing four qualitatively distinct regions of the QST landscape.

\textbf{High volume, low awareness} (lower-right; $n > 6$, awareness $< 23.0\%$): Test generation and debugging/repair constitute the most significant methodological gap in the current landscape. These dominant paradigms shape a large portion of QST research yet remain weakly aligned with noisy and hardware-constrained execution realities.

\textbf{High volume, high awareness} (upper-right; $n > 6$, awareness $\geq 23.0\%$): Oracle/Assertion-Based Testing (11 papers, 27.3\%) is the only paradigm currently occupying this region. Its position reflects the field's progress towards probabilistic semantics at the oracle level. However, no paradigm yet achieves \emph{majority} noise-awareness ($> 50\%$) at this volume, indicating that even the most hardware-aware dominant paradigm leaves the majority of its contributions implicitly idealized.

\textbf{Low volume, high awareness} (upper-left; $n \leq 6$, awareness $\geq 23.0\%$): Unit testing (6 papers, 50.0\%), the residual \emph{Other} category (6 papers, 50.0\%), and mutation testing (4 papers, 25.0\%) represent the most hardware-aware paradigms at their volume scale, although each rests on six or fewer papers.

\textbf{Low volume, low awareness} (lower-left; $n \leq 6$, awareness $< 23.0\%$): Coverage-based testing (5 papers, 0\%), metamorphic testing (2 papers, 0\%), and differential testing (2 papers, 0\%) are underexplored, with no noise-aware contributions at all. These paradigms represent both the smallest and the least hardware-aware regions of the landscape.

\subsubsection{Representative Frameworks by Paradigm}
\label{subsec:paradigm_synthesis}

\begin{table*}[htbp]
\caption{Paradigm-level synthesis of the analyzed corpus: size, noise-awareness rate, representative frameworks, and observed evaluation pattern of each paradigm. Bracketed values are Wilson 95\% confidence intervals.}
\label{tab:comparative_synthesis}
\centering
\footnotesize
\begin{tabular}{L{3.0cm} L{0.8cm} L{1.5cm} L{4.3cm} L{4.8cm}}
\toprule
\textbf{Paradigm} & \textbf{N} & \textbf{Noise-Aware \% [95\% CI]} & \textbf{Representative Frameworks (from analyzed corpus)} & \textbf{Observed Evaluation Pattern} \\
\midrule
Test Generation & 14 & 14.3\newline[4, 40] & QuanFuzz~\cite{paper_103}, QuSBT~\cite{paper_17}, Quantum Concolic Testing~\cite{paper_45}, From Ideal to Noisy~\cite{paper_94} & Predominantly simulator-oriented evaluation; the two noise-aware contributions are recent 2024--2025 additions. \\
Oracle / Assertion-Based Testing & 11 & 27.3\newline[10, 57] & Statistical Assertions~\cite{huang2019statistical}, Dynamic Runtime Assertions~\cite{paper_122}, Test Points for Online Monitoring~\cite{paper_15}, Commuting Pauli Strings~\cite{paper_39} & Methodologically diverse oracle-construction mechanisms, including probabilistic and assertion-based validation; third-highest noise-awareness rate among major paradigms. \\
Debugging / Repair / Fault Localization & 11 & 18.2\newline[5, 48] & HornBro~\cite{paper_44}, Framework for Debugging Quantum Programs~\cite{paper_54}, Mutation-based Fault Localization~\cite{paper_109}, Bloch Vector Assertions~\cite{paper_115} & Subset primarily focused on localization, repair, and debugging workflow support; only HornBro reports physical-hardware evaluation. \\
Unit Testing & 6 & 50.0\newline[19, 81] & QUT~\cite{paper_114}, Context-Aware Unit Testing~\cite{paper_117}, Multi-Subroutine Testing~\cite{paper_71} & Tied for the highest noise-awareness proportion; two of the three noise-aware papers report noisy-simulator evaluation. \\
Other & 6 & 50.0\newline[19, 81] & Understanding Quantum Noise~\cite{paper_67}, QuanTest~\cite{paper_36}, Distinguishing Bugs from Noise~\cite{paper_116} & Heterogeneous category including empirical noise-characterization, quantum-ML testing, and analytics-driven validation; tied with Unit Testing for the highest noise-awareness proportion. \\
Coverage-Based Testing & 5 & 0.0\newline[0, 43] & Quito~\cite{wang2021quito}, QuraTest~\cite{paper_6}, Gate Branch Coverage~\cite{paper_35} & Focused primarily on adequacy criteria and coverage metrics under idealized execution assumptions; zero hardware-realistic evaluation reported. \\
Mutation Testing & 4 & 25.0\newline[5, 70] & Muskit~\cite{paper_1}, QCRMut~\cite{paper_113}, Mutation Testing With Qiskit~\cite{paper_74}, Quantum Circuit Mutants~\cite{paper_78} & Primarily centered on mutation operators and mutation-analysis workflows; limited explicit engagement with noise-aware evaluation. \\
Metamorphic Testing & 2 & 0.0\newline[0, 66] & Metamorphic Testing of Oracle Programs~\cite{abreu2022metamorphic}, Asserting Shor Implementations~\cite{paper_27} & Small subset focused on metamorphic relations and oracle limitations; no hardware-realistic evaluation reported. \\
Differential Testing & 2 & 0.0\newline[0, 66] & Equivalence and Unitarity Checking~\cite{paper_70}, Black-box Testing Framework~\cite{long2025black} & Small subset focused on behavioral-equivalence validation and black-box testing strategies. \\
\midrule
\textbf{Total} & \textbf{61} & \textbf{23.0}\newline[14, 35] & -- & Ideal-simulator evaluation dominates; 11 papers (18.0\%) report hardware-realistic evaluation. \\
\bottomrule
\end{tabular}
\end{table*}

Table~\ref{tab:comparative_synthesis} summarizes representative frameworks and their reported evaluation configurations. We restrict the discussion to capabilities and configurations explicitly reported in each paper or confirmed through targeted manual inspection, and avoid inferring backend realism from terminology alone.

\paragraph*{Mutation testing} The mutation-testing subset contains four papers, of which one is classified as noise-aware. The subset is anchored by the mutation-analysis tool \emph{Muskit}~\cite{paper_1} and a Qiskit-focused case study of mutation operators~\cite{paper_74}, together with a circuit-mutant generator (\emph{QCRMut}~\cite{paper_113}) and a recent empirical analysis of quantum circuit mutants~\cite{paper_78}. Two of the four contributions~\cite{paper_74, paper_78} report evaluation exclusively on ideal simulators, \emph{Muskit} does not explicitly report its evaluation backend, and \emph{QCRMut} reports ideal-simulator evaluation alongside hardware-platform mentions without an explicit noisy-simulator configuration. The subset as a whole focuses on mutation operators, mutant-generation strategies, and mutation-analysis workflows for quantum circuits and quantum software artifacts; explicit engagement with hardware-realistic noise remains limited.

\paragraph*{Differential testing} The differential-testing subset contains two papers, both classified as noise-oblivious and both reporting evaluation exclusively on ideal simulators. The two contributions address equivalence, identity, and unitarity checking in black-box settings~\cite{paper_70, long2025black}. Neither paper explicitly addresses simulator-versus-hardware behavioral comparison, despite differential testing being naturally suited to such comparison. The paradigm therefore remains lightly represented in the corpus and has not yet been substantively extended to noisy quantum execution.

\paragraph*{Metamorphic testing} The metamorphic-testing subset contains two papers, both classified as noise-oblivious and both reporting evaluation on ideal simulators only. One contribution develops metamorphic relations for oracle quantum programs~\cite{abreu2022metamorphic}; the other applies metamorphic testing to assert the correctness of Shor's algorithm implementations~\cite{paper_27}. The subset illustrates ongoing interest in relational validation mechanisms for quantum software but has not yet incorporated hardware-realistic execution into its evaluation strategy.

\paragraph*{Coverage-based testing} Coverage-based testing contains five papers and no noise-aware classifications under the adopted operationalization. The subset includes the coverage-guided test generator \emph{Quito}~\cite{wang2021quito} together with four further contributions that develop quantum-specific coverage criteria~\cite{paper_6}, gate-branch coverage metrics~\cite{paper_35}, an assessment of input/output coverage criteria~\cite{paper_59}, and a coverage-guided framework for quantum neural networks~\cite{paper_107}. All five report evaluation on ideal simulators only. The subset's central focus is adequacy criteria, coverage metrics, and coverage-guided test generation under idealized execution settings.

\paragraph*{Unit testing of quantum subroutines} The unit-testing subset contains six papers, three of which are classified as noise-aware: \emph{QUT}~\cite{paper_114}, a context-aware unit-testing framework for quantum subroutines~\cite{paper_117}, and a multi-subroutine testing approach that progresses from unit to integration testing~\cite{paper_71}. Two of the three noise-aware contributions (\emph{QUT} and the context-aware framework) report evaluation on noisy simulators; the multi-subroutine work reports evaluation on an ideal simulator despite engaging with noise-related terminology at the methodological-framing level. The remaining three papers also evaluate on ideal simulators, and one of them, a recent study on the feasibility of quantum unit testing~\cite{paper_112}, explicitly scopes itself to fault-tolerant or noiseless execution. This polarization, in which two of six papers explicitly report noisy-simulator evaluation while the other four evaluate only on ideal simulators, distinguishes unit testing from the larger paradigms in the corpus, where ideal-simulator evaluation dominates.

\paragraph*{Oracle and assertion-based testing} The oracle/assertion subset contains eleven papers after manual-review integration. Representative contributions include the statistical-validation framework of Huang and Martonosi~\cite{huang2019statistical}, dynamic runtime assertion mechanisms~\cite{paper_122}, online monitoring through test points~\cite{paper_15}, and Pauli-string-based validation on IBM quantum hardware~\cite{paper_39}, yielding the paradigm's 27.3\% noise-awareness rate, the third-highest among the major paradigms. The subset exhibits methodological diversity in oracle construction, including assertion-based, probabilistic, and statistical-validation mechanisms.

\paragraph*{Debugging, repair, and fault localization} The debugging/repair subset contains eleven papers after manual-review integration. Two effects of that integration are worth noting. First, several papers originally matched on debugging-oriented keywords but whose primary contribution is oracle or assertion construction, notably the projection-based runtime assertion framework~\cite{li2020projection}, were reassigned out of debugging/repair into oracle/assertion-based testing under the oracle--debugging overlap rule. Second, a study evaluating mutation-based fault localization for quantum programs~\cite{paper_109} was reassigned out of mutation testing into debugging/repair on the grounds that its primary contribution is fault localization rather than mutation analysis.

Following inspection, the reviewer also retained several papers whose titles combine debugging-oriented and assertion-oriented terminology but whose primary focus remains debugging, namely the CUDA-Q statistical-assertion framework~\cite{li2025statistical}, the Bloch-vector assertion mechanism~\cite{paper_115}, and a statistical bug-locating method~\cite{sato2025bug}. Representative systems for the paradigm as a whole include the automated quantum-program repair framework \emph{HornBro}~\cite{paper_44}, a general debugging framework for quantum programs~\cite{paper_54}, and the mutation-based fault localization study~\cite{paper_109} mentioned above.

Two of the eleven papers are classified as noise-aware: \emph{HornBro}, which reports physical-hardware evaluation, and the Bloch-vector assertion work~\cite{paper_115}, which reports noisy-simulator evaluation. The resulting 18.2\% noise-awareness rate places the paradigm well below the corpus mean. Although several papers in this subset incorporate testing infrastructure internally, the primary methodological focus is bug localization, repair, debugging workflow support, or fault diagnosis.

\paragraph*{Test generation} Test generation remains the largest paradigm in the corpus with fourteen papers, two of which are classified as noise-aware: a property-based testing approach adapted to real-world noisy quantum computers~\cite{paper_94} and a dynamic random-testing strategy with hybrid distance metrics~\cite{paper_52}. The subset spans a wide range of generation strategies, including the fuzz-testing framework \emph{QuanFuzz}~\cite{paper_103}, the search-based testing approach \emph{QuSBT}~\cite{paper_17}, and the symbolic-execution framework \emph{Quantum Concolic Testing}~\cite{paper_45}, together with contributions on fuzz, combinatorial, property-based, random, and concolic execution.

Thirteen of the fourteen papers report evaluation on ideal simulators only; the property-based testing work~\cite{paper_94} is the sole contribution in this paradigm reporting evaluation on a noisy simulator, while the dynamic random-testing strategy~\cite{paper_52} engages with noise-related terminology (notably fidelity) at the methodological-framing level but reports evaluation on an ideal simulator. The two noise-aware contributions are recent 2024--2025 additions and represent the first explicit engagements with noisy execution within this paradigm.

\paragraph*{Other contributions} Six papers could not be reliably mapped onto one of the eight paradigms: an empirical characterization of quantum-noise impact on quantum programs~\cite{paper_67}, the entanglement-guided neural-network testing framework \emph{QuanTest}~\cite{paper_36}, a statistical approach for distinguishing software bugs from hardware noise~\cite{paper_116}, a testing framework for distributed quantum learning software (\emph{SQUAD})~\cite{paper_81}, an analytics-driven user-assistance system for quantum neural networks (\emph{AQUA})~\cite{paper_123}, and a runtime dynamic-analysis engine (\emph{EQuaTE})~\cite{paper_124}. The first three are noise-aware, yielding a 50.0\% rate that ties with unit testing.

\begin{answerbox}{Answer to RQ2}
Noise-awareness varies across paradigms but remains a minority concern in most of them. Unit testing and the residual \emph{Other} category reach 50.0\% (3 of 6 each), followed by oracle/assertion-based testing (27.3\%) and mutation testing (25.0\%). The two largest paradigms, test generation (14.3\%) and debugging/repair (18.2\%), fall below the corpus rate, and coverage-based (0/5), metamorphic (0/2), and differential (0/2) testing contain no noise-aware contributions. Given the small subsamples, these contrasts are descriptive rather than statistically established (Freeman--Halton $p \approx 0.40$).
\end{answerbox}

\subsection{RQ3: Evolution over Time}
\label{subsec:temporal_noise}

\begin{figure}[htbp]
\centering
\includegraphics[width=\columnwidth]{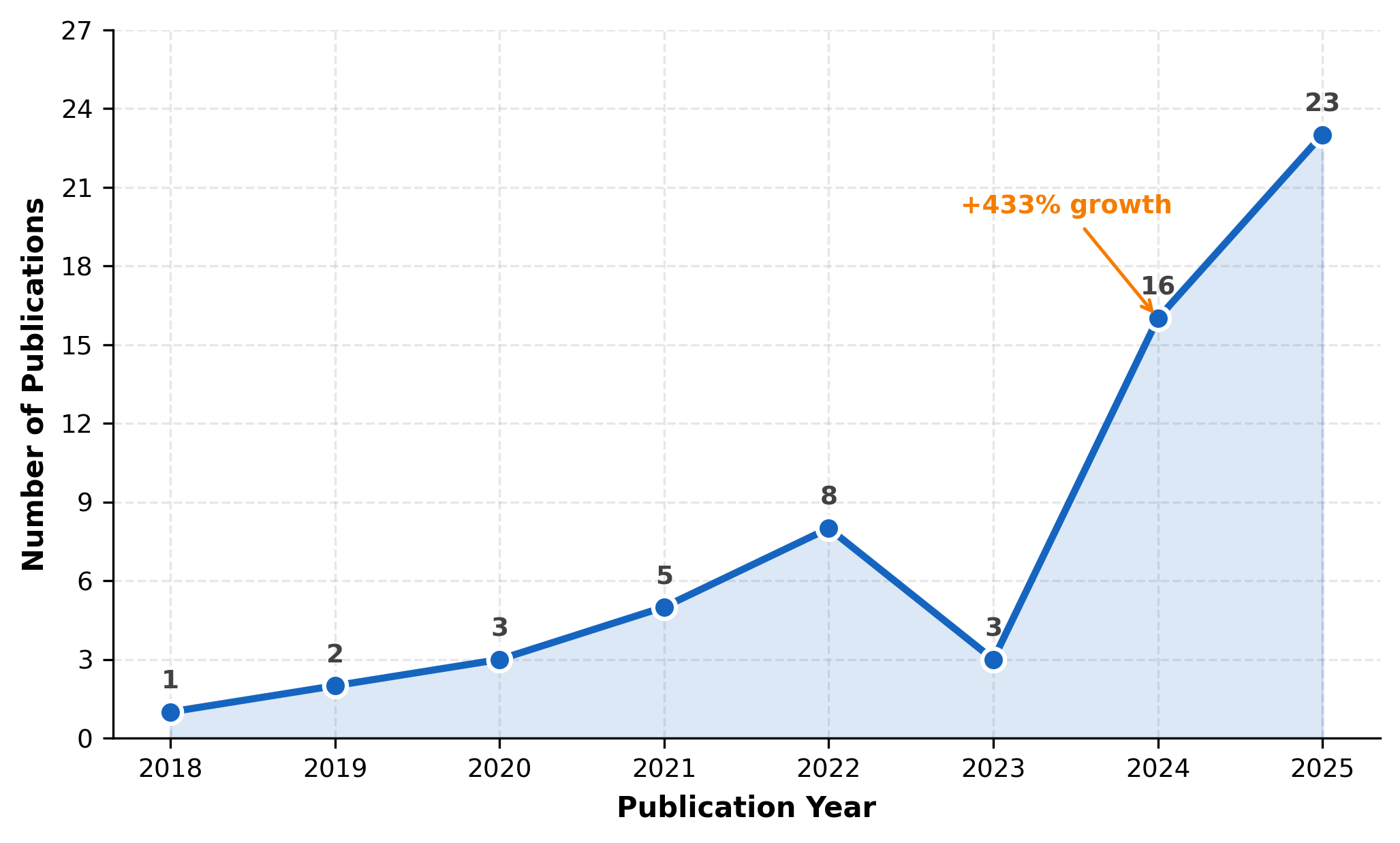}
\caption{Temporal evolution of quantum software testing publications (2018--2025). The line chart illustrates rapid growth in research activity, with pronounced acceleration beginning in 2024 (a 433\% increase over 2023).}
\label{fig:publication_growth}
\end{figure}

QST publication volume grew from a single paper in 2018 to 23 in 2025: 1--5 papers annually between 2018 and 2021, 8 in 2022, a temporary contraction to 3 in 2023, and rapid expansion thereafter, with 16 papers in 2024 (a 433.3\% increase over 2023) and a further 43.8\% increase in 2025 (Figure~\ref{fig:publication_growth}). The question is whether noise-awareness kept pace with this growth.

\begin{figure}[htbp]
\centering
\includegraphics[width=\columnwidth]{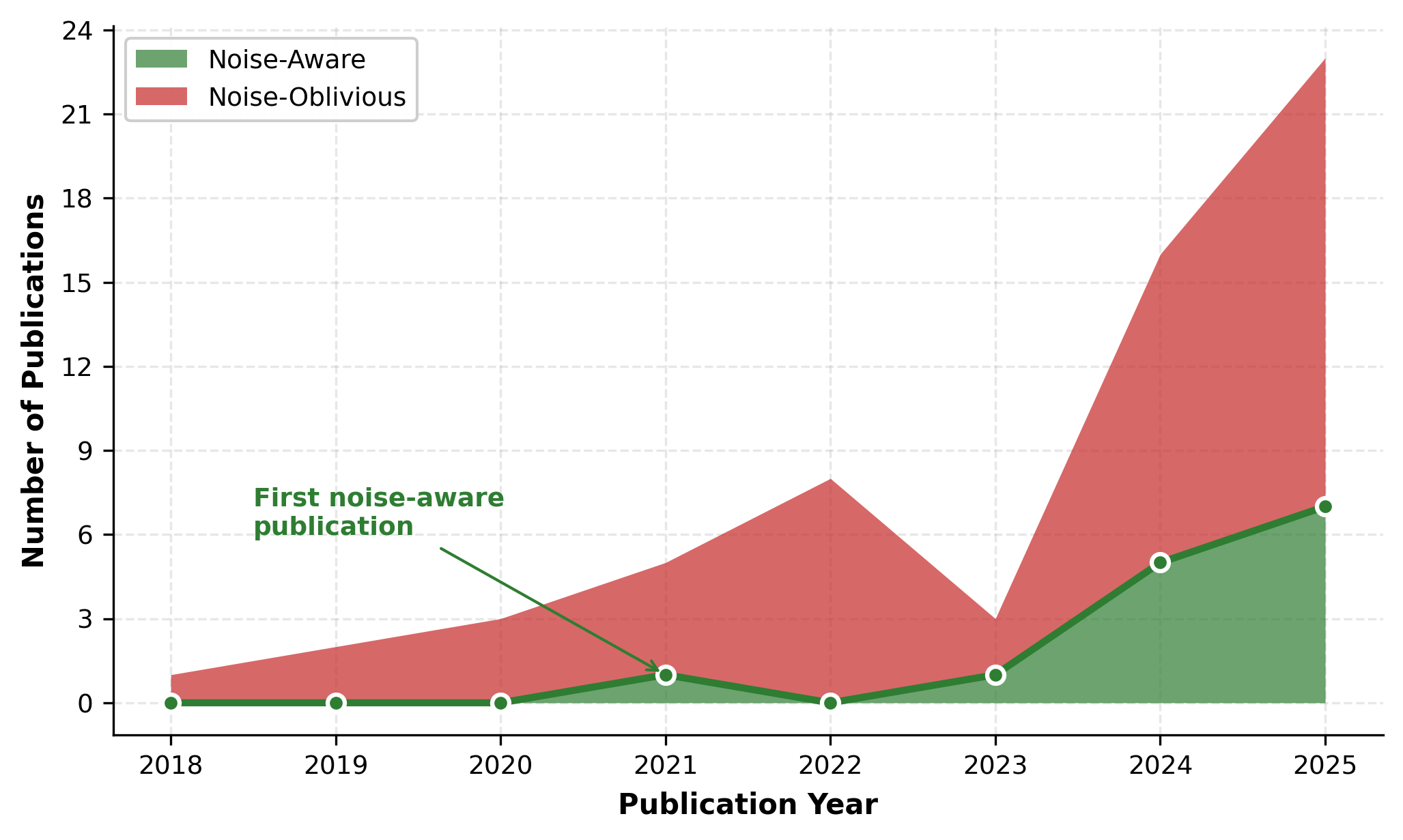}
\caption{Temporal evolution of noise-awareness in quantum software testing research shown as a stacked area chart. Noise-aware papers (green) increase gradually after 2021 but remain a minority through 2025. The absolute gap between noise-oblivious and noise-aware work fluctuates between years and is widest in 2025.}
\label{fig:temporal_evolution}
\end{figure}

Figure~\ref{fig:temporal_evolution} depicts the temporal evolution of noise-awareness using a stacked area chart, revealing three distinct periods. The \emph{early period} (2018--2020) is characterized by complete noise-obliviousness, with zero noise-aware papers among six publications (0.0\%). The \emph{emergence period} (2021--2022) begins with the first noise-aware contribution in 2021 (1 out of 5 papers, 20.0\%), followed by a regression to 0\% noise-awareness in 2022 despite increased publication volume. The \emph{maturation period} (2023--2025) shows sustained but incomplete progress, with 13 noise-aware papers among 42 publications (31.0\%).

Noise-aware papers increased from 1 (33.3\%) in 2023 to 5 (31.2\%) in 2024 and 7 (30.4\%) in 2025. Nevertheless, even in the most recent year, noise-oblivious papers outnumber noise-aware papers by approximately 2.3:1 (16 to 7), underscoring the persistence of idealized assumptions despite widespread recognition of NISQ limitations. Moreover, the absolute gap between noise-oblivious and noise-aware work, although it fluctuates between years (1, 2, 3, 3, 8, 1, 6, and 9 papers from 2018 to 2025), is widest in the most recent year.

A caveat applies to the lexical reading of this trend. The vocabulary on which the noise-awareness classifier operates, most notably the terms \emph{NISQ} and \emph{noise}, itself diffused through the quantum-computing literature over the same period, becoming common only from around 2018 onward. Part of the apparent rise in lexically noise-aware work may therefore reflect adoption of this terminology rather than a substantive methodological shift. We mitigate this confound by anchoring the temporal claim on the evaluation-realism series below, which depends on the execution backends each paper reports rather than on abstract vocabulary and is correspondingly less sensitive to lexical fashion.

\paragraph*{Evaluation realism over time} Across the three periods identified above, the proportion of papers explicitly reporting evaluation on either physical hardware or a noisy simulator is: 1 of 6 (16.7\%) in the early period (2018--2020), 1 of 13 (7.7\%) in the emergence period (2021--2022), and 9 of 42 (21.4\%) in the maturation period (2023--2025). The hardware-realistic evaluation rate is therefore highest in the most recent period, but it remains well below the rate at which idealized-simulator evaluation is reported, and the 2024--2025 cohort still relies predominantly on ideal-simulator validation.

\paragraph*{Trend test} To assess whether these changes amount to a monotonic trend, we applied a Cochran--Armitage test for trend across the eight annual cohorts, using publication year as the score. The share of noise-aware papers shows an increasing tendency that falls just short of conventional significance ($z = 1.95$, two-sided $p \approx 0.051$; exact permutation $p \approx 0.06$), and the share of papers reporting hardware-realistic evaluation shows no significant trend ($z = 0.90$, $p \approx 0.37$). The temporal pattern should therefore be read descriptively: noise-aware work appears late and grows in absolute number, but its share of the literature has not yet increased to a statistically detectable degree.

A notable concentration of hardware-realistic evaluation appears in the 2025 cohort. Six 2025 papers explicitly report evaluation under noise-realistic conditions: the unit-testing frameworks \emph{QUT}~\cite{paper_114} and a context-aware unit-testing extension~\cite{paper_117} (both on noisy simulators), the Bloch-vector assertion mechanism~\cite{paper_115} (noisy simulator), the \emph{HornBro} program-repair framework~\cite{paper_44} (physical hardware), the noise-adapted property-based testing work of Pontolillo et al.~\cite{paper_94} (noisy simulator), and a statistical bug-locating method~\cite{sato2025bug} (noisy simulator). This concentration is consistent with the interpretation that hardware-aware methodology is most actively developed in recently emerging QST subareas, particularly unit testing, assertion-based testing, and statistical debugging, rather than in established paradigms.

\begin{answerbox}{Answer to RQ3}
Noise-awareness emerged late and has grown in absolute number rather than transformatively. No paper from 2018--2020 is noise-aware; the first appears in 2021, and the 2023--2025 period reaches 31.0\% (13 of 42). Noise-oblivious work nevertheless dominates every year, and the absolute gap widened to 9 papers in 2025. Neither the noise-aware share (Cochran--Armitage $p \approx 0.051$) nor the hardware-realistic share ($p \approx 0.37$) shows a statistically significant trend, so the evidence indicates late, gradual emergence rather than a field-wide methodological transition.
\end{answerbox}

\subsection{RQ4: Remaining Gaps}
\label{subsec:rq4}

\subsubsection{Empirical Practice Patterns}
\label{subsec:practice_patterns}

Beyond how contributions are framed (RQ1, RQ2) and where they are evaluated (Section~\ref{subsec:evaluation_realism}), a third reading examines what each paper explicitly reports about its experimental setup: the oracle, inferential statistical tests, shot budget, benchmarks, and reproducibility artifacts (Table~\ref{tab:practice_patterns}). None of the differences between the noise-aware and noise-oblivious subsets in Table~\ref{tab:practice_patterns} is statistically significant (Fisher's exact $p \geq 0.36$ for every row), so the reporting profile is best read as field-wide rather than specific to either subset.

\begin{table}[htbp]
\caption{Explicit reporting of experimental-setup practices, overall and by noise-awareness classification. A practice is counted only when the paper states it explicitly.}
\label{tab:practice_patterns}
\centering
\footnotesize
\resizebox{\columnwidth}{!}{%
\begin{tabular}{l c c c}
\toprule
\textbf{Practice dimension}  & \textbf{Reported} & \textbf{Noise-Aware} & \textbf{Noise-Oblivious} \\
& \textbf{(of 61)}  & \textbf{(of 14)}     & \textbf{(of 47)} \\
\midrule
Test-oracle name             & 37 (60.7\%)       & 10 (71.4\%)          & 27 (57.4\%) \\
Inferential statistical test & 10 (16.4\%)       & 2 (14.3\%)           & 8 (17.0\%) \\
Shot budget                  & 35 (57.4\%)       & 10 (71.4\%)          & 25 (53.2\%) \\
Backend configuration        & 54 (88.5\%)       & 14 (100\%)           & 40 (85.1\%) \\
Artifact link                & 34 (55.7\%)       & 9 (64.3\%)           & 25 (53.2\%) \\
GitHub-style repository      & 26 (42.6\%)       & 6 (42.9\%)           & 20 (42.6\%) \\
\bottomrule
\end{tabular}}
\end{table}

\paragraph*{Test-oracle landscape} Thirty-seven papers (60.7\%) name an explicit test oracle; the remaining twenty-four leave the oracle implicit, inferable only from informal descriptions of the validation procedure. Among the named oracles, classified according to the oracle taxonomy of Li et al.~\cite{li2026methodological}, two recur with sufficient frequency to constitute de facto community practice, consistent with their observation that these are the most commonly employed oracles: the wrong-output oracle (WOO, 16 papers, 12 of which check the original program outcome) and the output-probability oracle (OPO, 23 papers), most often paired with Pearson's chi-squared test (16 papers across two spelling variants in the underlying data). Property-based oracles (PBO) appear in 11 papers, quantum-state oracles in 5 (three of which use quantum fidelity), and the dominant-output oracle (DOO) in 3. The distribution is therefore long-tailed: the two dominant oracle constructions together appear in 26 of the 37 papers that report an oracle at all, but a non-trivial 24 papers (39.3\% of the corpus) do not document this most basic methodological choice.

\paragraph*{Statistical-evaluation practice} Only 10 of 61 papers (16.4\%) accompany their experimental findings with at least one inferential statistical test. These ten span several test-generation contributions, including \emph{QuCAT}~\cite{paper_5}, \emph{QuSBT}~\cite{paper_17}, a search-driven failing-test-suite generator~\cite{paper_21}, a strategy of dynamic random testing with hybrid distance metrics~\cite{paper_52}, an application of combinatorial testing to quantum programs~\cite{paper_60}, a mutation-based test generator~\cite{paper_20}, and a specification-reduction approach for faster quantum software testing~\cite{paper_43}, together with a machine-learning-based noise-mitigation framework~\cite{paper_73}, an evaluation of mutation-based fault localization for quantum programs~\cite{paper_109}, and a study of output validation in quantum program testing~\cite{ye2025measurement}.

The Mann--Whitney $U$ test is the most frequently reported, appearing in 8 of the 10 papers, and is paired with Vargha--Delaney $\hat{A}_{12}$ effect sizes in 7 of those 8 papers. Fisher's exact, Kruskal--Wallis, Spearman's rank correlation, Wilcoxon signed-rank, and Cliff's $\delta$ are each reported in one or two papers as supplementary tests, and one paper reports a Kolmogorov--Smirnov test. The remaining 51 papers either do not report statistical tests or do not specify them. The reporting rate is not concentrated in the noise-oblivious subset: 2 of 14 noise-aware papers (14.3\%) and 8 of 47 noise-oblivious papers (17.0\%) report inferential tests. Probabilistic NISQ execution is widely acknowledged in the framing of QST research but is not, on the whole, accompanied by the inferential statistical machinery that such execution makes necessary.

\paragraph*{Shot-budget practice} Thirty-five papers (57.4\%) explicitly report a measurement shot budget. Of these, the dominant pattern is a fixed shot count (17 papers, 27.9\% of the corpus), with adaptive schemes (9 papers, 14.8\%), explicitly varied budgets (7 papers, 11.5\%), and mixed fixed/adaptive strategies (2 papers, 3.3\%) constituting the remainder. The 26 papers (42.6\%) that do not specify a shot budget include 4 noise-aware contributions, indicating that even within the noise-aware subset the most basic statistical-sampling parameter of NISQ execution is under-documented in nearly one third of cases (4 of 14, 28.6\%). Among reporting papers, fixed values cluster around powers of two associated with backend defaults (1024, 8192) or round numbers (100, 2000, 3000); a small number of contributions use adaptive schemes that scale with the number of possible outputs or with the rank of the program specification.

\paragraph*{Benchmark concentration} The benchmark distribution exhibits strong concentration on textbook quantum algorithms. The most frequently used benchmarks across the corpus are the Quantum Fourier Transform, including its inverse (30 papers), Grover search (16), Bernstein--Vazirani (15), Quantum Phase Estimation (15), Simon's algorithm (9), Quantum Teleportation (9), Deutsch--Jozsa (9), and Shor's algorithm (6), consistent with Li et al.~\cite{li2026methodological}, who also identify the Quantum Fourier Transform as the most frequent program type. Several papers additionally rely on artificial programs (8 papers), 8 papers do not specify their benchmark programs, and only a small number draw on established benchmark repositories such as Bugs4Q (6 papers), MQT Bench (4), VeriQBench (2), or RevLib (1). The concentration is methodologically consequential: the same handful of textbook circuits is reused across nearly all subsets of the corpus, including the noise-aware subset, raising the question of whether observed noise-related findings would generalize to the more application-oriented programs that NISQ-era practice increasingly involves.

\paragraph*{Reproducibility patterns} Thirty-four papers (55.7\%) provide an explicit artifact link. Of these, 26 (42.6\% of the corpus) link to GitHub (25 to a repository and one to a GitHub-hosted project page), while 12 deposit artifacts on long-term archives (8 on Zenodo, 3 on Figshare, and 1 on OSF); 4 papers provide both. The substantial difference between artifact availability via non-archival repositories and via archival deposit is itself a reproducibility signal: GitHub repositories are widely used but are not guaranteed to persist for the lifetime of a published scientific record. Artifact links are modestly more frequent in the noise-aware subset (64.3\% versus 53.2\%), while GitHub-style repository links are nearly identical (42.9\% versus 42.6\%), indicating that artifact-release practice in QST research has not yet specialized substantially to the hardware-aware sub-community.

\subsubsection{Synthesis of Gaps}
\label{subsec:gaps}

Combining the lexical, evaluation-realism, and practice-pattern evidence surfaces five interlocking gaps, ordered from the most structural to the most operational.

\emph{Gap~1 (structural).} No QST paradigm currently combines broad adoption with majority noise-awareness. The high-volume, high-awareness region of Figure~\ref{fig:volume_vs_awareness} (Section~\ref{subsec:synthesis}) is occupied by a single paradigm (Oracle/Assertion-Based Testing at 27.3\%), and no paradigm yet exceeds the 50\% majority threshold at $n>6$. A mature, widely adopted hardware-aware QST methodology has not yet emerged.

\emph{Gap~2 (dominant paradigms).} Test generation (14 papers) and debugging/repair (11 papers), which together account for 25 of 61 contributions (41.0\%), exhibit noise-awareness rates of 14.3\% and 18.2\% respectively, both below the corpus mean of 23.0\%. The paradigms shaping the bulk of QST research engage with hardware noise less than the field as a whole, indicating that methodological dominance and hardware-conscious testing practice remain weakly aligned.

\emph{Gap~3 (silent paradigms).} Coverage-based testing (0/5), metamorphic testing (0/2), and differential testing (0/2) contain no noise-aware contributions in the analyzed corpus. The coverage-based result is methodologically substantive, indicating that adequacy criteria proposed for quantum programs have not been adapted to stochastic execution semantics. The metamorphic and differential results rest on smaller subsets that limit individual interpretation but, taken collectively, suggest that relational and cross-platform validation mechanisms have not yet been extended to NISQ execution.

\emph{Gap~4 (evaluation realism).} Only 18.0\% of papers explicitly report evaluation on hardware-realistic backends, and 11.5\% leave the backend configuration unspecified (Section~\ref{subsec:evaluation_realism}). Within the lexically noise-aware subset, 5 of 14 papers (35.7\%) frame themselves around noise but evaluate exclusively on ideal simulators. Although the two dimensions are strongly associated overall (Section~\ref{subsec:evaluation_realism}), they diverge for a substantial fraction of the noise-aware subset.

\emph{Gap~5 (practice-pattern).} The methodological elements most directly required by probabilistic NISQ execution are also the least consistently documented in the corpus: explicit test oracles are named in 60.7\% of papers, shot budgets in 57.4\%, and inferential statistical tests in only 16.4\% (Section~\ref{subsec:practice_patterns}). Noise-aware papers name an oracle (71.4\% versus 57.4\%), report a shot budget (71.4\% versus 53.2\%), and provide an artifact link (64.3\% versus 53.2\%) more often than noise-oblivious papers, yet 4 of the 14 noise-aware papers name no oracle and 4 specify no shot budget, and inferential statistical tests are reported at similarly low rates in both subsets (14.3\% versus 17.0\%). None of these subset differences is statistically significant (Fisher's exact $p \geq 0.36$). Reporting-level support for probabilistic evaluation therefore remains incomplete across the field, including within the hardware-aware sub-community.

These five gaps are mutually reinforcing rather than independent: the structural gap is sustained by the under-engagement of the dominant paradigms (Gap~2) and by the silence of the adequacy/relational/cross-platform paradigms (Gap~3); the practice-pattern gap (Gap~5) undermines the reproducibility and empirical force of even the noise-aware contributions that do exist; and the divergence between framing and evaluation (Gap~4) shows that lexical engagement with noise does not, on its own, yield hardware-realistic validation. Section~\ref{subsec:disc_agenda} maps each gap to a concrete research direction.

\begin{answerbox}{Answer to RQ4}
Five gaps remain: (1)~no paradigm combines broad adoption with majority noise-awareness; (2)~the dominant paradigms, test generation and debugging/repair (41.0\% of the corpus), are less noise-aware than the field as a whole; (3)~coverage-based, metamorphic, and differential testing have not engaged with noise; (4)~noise-aware framing is often not matched by hardware-realistic evaluation; and (5)~oracles, shot budgets, and inferential statistics are under-reported even within the noise-aware subset.
\end{answerbox}

\section{Discussion}
\label{sec:discussion}

This study combines two complementary lines of evidence: a deterministic lexical classification of how QST contributions are framed (Sections~\ref{subsec:overall_distribution}--\ref{subsec:paradigm_synthesis}), and the evaluation configurations and reporting practices that the same contributions explicitly document (Sections~\ref{subsec:evaluation_realism} and~\ref{subsec:practice_patterns}). Both indicate that noise-awareness remains a minority concern and that the field is in a transitional phase between simulator-centric methodological development and fully hardware-conscious testing practice.

More fundamentally, the findings suggest that a substantial portion of contemporary QST methodology remains only partially aligned with the execution semantics of NISQ-era quantum hardware. Many dominant testing paradigms continue to implicitly inherit assumptions from classical software testing, including deterministic execution behavior, stable oracles, and hardware-independent validation. However, practical quantum execution is probabilistic, backend-sensitive, and strongly shaped by calibration and noise characteristics. The resulting gap is therefore not merely a lack of explicit ``noise-aware'' terminology, but a broader methodological mismatch between prevailing QST abstractions and the physical execution environments for which quantum software must increasingly be validated.

The empirical findings of Section~\ref{sec:empirical-landscape} admit several interpretive threads. This section synthesizes the results into broader methodological observations and discusses implications for future research on oracle construction, adequacy modeling, probabilistic execution, and hardware-aware testing design.

\subsection{Interpreting the Noise-Awareness Gap}
\label{subsec:disc_gap}

The most consequential pattern observed in the analyzed corpus is not merely the overall 77.0\% noise-oblivious rate, but the structural imbalance between methodological maturity and noise-awareness across testing paradigms. The two largest paradigms in the corpus, test generation and debugging/repair/fault localization, remain overwhelmingly noise-oblivious despite their methodological prominence. Conversely, smaller and more recent paradigms such as unit testing exhibit substantially higher noise-awareness rates. This apparent inverse relationship between paradigm dominance and noise-awareness, although not statistically established at this sample size (Section~\ref{subsec:by_technique}), suggests that many influential QST methodologies continue to inherit assumptions from classical software testing, where deterministic execution and reproducible behavior are natural expectations.

\subsubsection*{Why dominant paradigms remain idealized}

Test generation and debugging both possess strong methodological roots in classical software engineering. Classical test generation techniques generally assume deterministic program behavior, while classical debugging methodologies typically rely on reproducible failures and stable execution traces. Quantum adaptations of these paradigms therefore emerge most naturally under idealized execution assumptions, where nondeterministic hardware effects can be abstracted away through simulation.

This inheritance effect appears throughout the analyzed corpus. Many representative frameworks focus primarily on their core methodological contribution, for example, test-input generation, symbolic exploration, fault localization, or repair strategies, while treating noise-handling as peripheral to the central research objective. Consequently, hardware-noise integration is frequently deferred, simplified, or excluded from the primary evaluation setting.

\subsubsection*{Why unit testing appears more hardware-aware}

The comparatively high noise-awareness rate observed in the unit testing subset (50.0\%, 3 of 6 papers) is statistically modest because of the subset size, but methodologically suggestive. Subroutine-level testing is naturally aligned with the scale at which contemporary quantum engineering is often performed: researchers typically reason about reusable circuit components, ansatz blocks, oracle implementations, and modular quantum subroutines rather than monolithic applications.

At this granularity, practical interaction with noisy execution becomes more tractable. Smaller quantum modules are often short enough that their behavior can still be meaningfully evaluated under NISQ constraints before noise overwhelms signal quality. Additionally, many unit-testing contributions emerged relatively recently, during a period in which hardware realism and noisy execution had already become central concerns in quantum computing research. The resulting methodologies therefore appear more tightly coupled to contemporary NISQ-era engineering practice than several older testing paradigms.

\subsubsection*{Why noise-awareness emerged late}

The late and gradual emergence of noise-awareness documented in Section~\ref{subsec:temporal_noise} is consistent with the inheritance effect described above: early QST work transferred classical paradigms into the quantum setting before hardware access and noise characterization became routine. The noise-aware contributions that did emerge cluster in recent work on unit testing, noise-mitigation-aware testing, and hardware-oriented validation, which suggests gradual methodological adaptation originating in newer subareas rather than a field-wide transition toward hardware-aware testing practice.

\subsection{Structural Reading of the QST Landscape}
\label{subsec:disc_structure}

The volume-versus-awareness view (Section~\ref{subsec:synthesis}) shows that the methodological region corresponding to simultaneously mature and hardware-conscious QST practice remains largely unoccupied, and that noise-oblivious assumptions are concentrated in the field's dominant paradigms (Gaps~1 and~2 in Section~\ref{subsec:gaps}). Because these paradigms strongly shape the methodological direction of QST research, their current abstraction level may inadvertently reinforce the continued prioritization of idealized execution models within the broader research ecosystem.

This observation highlights a central opportunity for future research. One possible trajectory involves the maturation of currently noise-aware paradigms into broader methodological adoption. Another involves the redesign of dominant paradigms such as test generation and debugging around probabilistic and hardware-constrained execution semantics rather than idealized simulation assumptions.

The paradigm-level synthesis further suggests that many noise-aware contributions currently exist as isolated systems rather than as components of mature methodological subfields. Representative recent contributions such as HornBro~\cite{paper_44}, QUT~\cite{paper_114}, and Bloch Vector Assertions~\cite{paper_115} illustrate growing interest in hardware-conscious testing and validation, but these efforts have not yet collectively matured into dominant research paradigms within the broader QST landscape.

\subsection{Engineering and Methodological Implications}
\label{subsec:disc_implications}

The findings carry implications across several methodological subareas of QST, including oracle construction, adequacy modeling, differential validation, and hardware-aware software engineering.

\subsubsection*{Oracle construction}

Oracle and assertion-based testing exhibits one of the highest noise-awareness rates among the major paradigms in the analyzed corpus. This result is methodologically intuitive because oracle construction is closely tied to probabilistic reasoning, statistical validation, and interpretation of measurement outcomes. Assertions over quantum outputs frequently require reasoning about distributions, tolerances, repeated execution, and confidence under nondeterministic behavior.

The analyzed oracle-oriented contributions exhibit substantial methodological diversity, including projection-based assertions, statistical assertions, runtime monitoring, contract-based validation, and probabilistic output reasoning. Nevertheless, manual review indicates that several papers employing oracle-oriented terminology are fundamentally debugging-oriented in their primary methodological contribution. This observation reinforces the importance of ambiguity-aware classification and semantic review within QST mapping studies.

More broadly, the results suggest that reusable and noise-conscious oracle mechanisms remain an open research challenge. In particular, principled oracle constructions that incorporate statistical confidence, uncertainty tolerance, and hardware-noise characteristics as first-class testing parameters could significantly strengthen the practical applicability of assertion-based QST methodologies.

\subsubsection*{Adequacy modeling}

Coverage-based testing exhibits no noise-aware contributions in the analyzed corpus. This result is methodologically significant because adequacy criteria determine how testing completeness is formalized within a paradigm.

Classical adequacy notions such as statement, branch, and path coverage assume deterministic execution and discrete execution events. Quantum execution, however, is fundamentally probabilistic and measurement-dependent. Consequently, noise-aware adequacy modeling for quantum programs cannot simply reuse classical definitions without modification.

The absence of noise-aware adequacy criteria in the analyzed literature therefore highlights a substantial methodological gap. Future work may require probabilistic coverage notions, distribution-aware adequacy metrics, information-theoretic coverage measures, or execution-budget-aware adequacy formalizations capable of operating under noisy quantum execution.

\subsubsection*{Differential validation and cross-platform testing}

Differential testing remains only lightly represented in the analyzed corpus despite its natural relevance to simulator-versus-hardware validation and behavioral-equivalence analysis. The two identified differential-testing contributions focus primarily on black-box validation and equivalence-oriented reasoning rather than explicit hardware-noise characterization.

This limited representation is noteworthy because differential validation provides a natural mechanism for detecting behavioral mismatches between simulators, noisy execution environments, and physical devices. Programs that appear correct under idealized simulation may exhibit substantially different behavior under realistic execution constraints. A mature differential-testing infrastructure could therefore become an important methodological foundation for future hardware-aware QST practice.

\subsubsection*{Hardware-aware software engineering}

A broader implication concerns the relationship between testing methodologies and the execution environments on which quantum software ultimately operates. The dominance of noise-oblivious methodologies suggests that much of the current QST literature still treats hardware effects as external complications rather than as structural aspects of software quality engineering.

In mature classical engineering domains, testing methodologies for safety-critical and reliability-sensitive systems frequently incorporate hardware behavior directly into validation and fault analysis processes. The analyzed QST literature does not yet exhibit a similarly established hardware-conscious testing tradition. Developing such a tradition will likely require not only new testing techniques, but also a broader methodological shift toward treating hardware-aware evaluation as a central research objective rather than an optional extension.

The findings further suggest that hardware realism should increasingly be treated as a first-class evaluation dimension within QST research. At present, many methodologies are validated primarily through idealized simulator-based execution, even when the claimed testing objectives concern behavior fundamentally shaped by noisy hardware dynamics. This creates a risk of methodological overgeneralization: testing techniques validated under ideal execution assumptions may not preserve their effectiveness, adequacy properties, or fault-detection capabilities under realistic NISQ conditions. Systematic characterization of backend realism, noise models, transpilation effects, and shot-budget sensitivity may therefore become an essential component of mature empirical QST evaluation practice.

\subsection{Recommendations for Future QST Research}
\label{subsec:recommendations}
\label{subsec:disc_agenda}

The findings support several concrete recommendations for future QST research. Table~\ref{tab:agenda} maps each gap identified in Section~\ref{subsec:gaps} to the corresponding recommendation.

\begin{table}[htbp]
\caption{Identified gaps (Section~\ref{subsec:gaps}) and the corresponding research directions.}
\label{tab:agenda}
\centering
\footnotesize
\begin{tabular}{L{3.1cm} L{4.75cm}}
\toprule
\textbf{Gap} & \textbf{Research direction} \\
\midrule
Gap~1: No high-volume, majority noise-aware paradigm & R5: Longitudinal monitoring against the volume-versus-awareness baseline \\
Gap~2: Dominant paradigms under-engage with noise & R1: Re-ground test generation and debugging in probabilistic, hardware-constrained semantics \\
Gap~3: Silent adequacy, relational, and differential paradigms & R3: Noise-aware adequacy and metamorphic relations; R4: cross-platform differential testing \\
Gap~4: Framing without hardware-realistic evaluation & R4: Differential validation across ideal simulators, noisy simulators, and devices \\
Gap~5: Under-reported oracles, shots, and statistics & R2: Reusable probabilistic oracles and reporting standards \\
\bottomrule
\end{tabular}
\end{table}

\paragraph*{R1: Integrate noise-awareness into dominant paradigms \textnormal{(addresses Gap~2)}} Future work in test generation, debugging, repair, and related high-volume paradigms should explicitly investigate how probabilistic execution and hardware-noise behavior affect existing testing assumptions, fault models, and validation strategies.

\paragraph*{R2: Develop reusable probabilistic oracle mechanisms \textnormal{(addresses Gap~5)}} Future assertion-based and oracle-oriented methodologies should increasingly incorporate statistical confidence, uncertainty tolerance, and noise-aware reasoning directly into oracle construction and evaluation design. The under-documentation of oracle choices and inferential statistics in the empirical-practice profile (Section~\ref{subsec:practice_patterns}) makes this recommendation operationally concrete: improving reporting practice is a prerequisite for cumulative methodological progress.

\paragraph*{R3: Prioritize noise-aware adequacy modeling \textnormal{(addresses Gap~3)}} Coverage-based and metamorphic testing remain largely unexplored from a noise-aware perspective. These paradigms provide promising foundations for developing probabilistic adequacy criteria, distribution-aware coverage models, and relational validation mechanisms under noisy execution.

\paragraph*{R4: Expand differential and cross-platform testing \textnormal{(addresses Gaps~3 and~4)}} Differential validation infrastructures capable of comparing ideal simulators, noisy simulators, and physical devices could provide a principled mechanism for identifying execution mismatches and hardware-specific behavioral divergence, and would directly narrow the gap between lexically noise-aware framing and hardware-realistic evaluation documented in Section~\ref{subsec:evaluation_realism}.

\paragraph*{R5: Maintain longitudinal empirical monitoring of the QST ecosystem \textnormal{(addresses Gap~1)}} The deterministic classification framework developed in this study can be reapplied to future publication cohorts to track the evolution of testing paradigms, ambiguity patterns, noise-awareness integration, and hardware-conscious methodological practice over time. The volume-versus-awareness view of Section~\ref{subsec:synthesis} provides a structural baseline against which the eventual emergence of a high-volume, high-awareness paradigm can be empirically detected.

Taken together, the findings suggest that noise-awareness should not be viewed as a specialized subtopic of quantum software testing, but rather as a foundational property of testing methodologies intended for realistic quantum execution environments. As quantum software increasingly transitions from idealized simulation to deployment on noisy hardware platforms, the ability of QST methodologies to reason explicitly about probabilistic execution, backend variability, and hardware-induced uncertainty will likely become a defining requirement for the next generation of quantum software engineering research.

\section{Threats to Validity}
\label{sec:threats}

We discuss threats to validity following established empirical software engineering practice.

\subsection{Construct Validity}

Construct validity concerns the extent to which the operational criteria adopted in this study accurately capture the underlying concepts of \emph{testing paradigm} and \emph{noise-awareness}. The principal threat is that lexical classification may fail to identify papers employing non-canonical terminology or unconventional methodological framing. A paper may engage substantively with noisy quantum execution without explicitly using the keywords included in our framework, while another paper may mention canonical testing or noise-related terminology only peripherally.

We mitigated this threat by deriving the lexicons from both classical and quantum testing terminology and refining them through ambiguity analysis, by distinguishing strong, weak, hardware-context, and idealization indicators (Section~\ref{subsec:noise_operationalization}), by applying the idealization-override rule so that hardware-platform mentions do not inflate noise-awareness in explicitly idealized work, and by using word-boundary matching to suppress spurious substring matches. The strong association between the abstract-level noise label and the evaluation backends extracted from the full papers (odds ratio 40.5, $p < 0.001$; Section~\ref{subsec:evaluation_realism}) provides convergent evidence for the construct.

A further construct-validity concern is that lexical matching does not directly capture semantic intent. Several papers in the QST literature employ terminology spanning multiple paradigms, particularly oracle/assertion terminology embedded within primarily debugging-oriented contributions. To mitigate this issue, the framework incorporates ambiguity-aware review mechanisms, including overlap flagging, semantic inspection, and manual reassignment for flagged cases. Thirteen papers entered the testing-category review queue under one or more ambiguity conditions; five papers were reassigned after structured manual inspection. This process reduces, but does not eliminate, the risk that methodological identity may be imperfectly represented through lexical evidence alone.

Two further construct-validity limitations follow from the design of the lexical pipeline. First, the keyword lexicons were refined through ambiguity analysis on a held-out subset of the corpus (Table~\ref{tab:testing_keywords}), so they are not fully independent of the data they classify; the blinded inter-rater validation (Section~\ref{subsubsec:interrater_validation}) provides an external check on the resulting classifications. Second, matching is lexical: apart from idealization markers, the pipeline does not distinguish affirmative from negated or contrastive mentions of noise terms, so a paper that mentions noise only to exclude it may be classified as noise-aware unless an idealization marker is present or the paper is routed to manual review.

Finally, the operationalization of noise-awareness itself constitutes a construct-validity limitation. The framework classifies papers according to explicit methodological engagement with noise-related execution concerns rather than according to actual experimental robustness or hardware realism. Consequently, a paper classified as noise-oblivious may still contain limited hardware experimentation, while a paper classified as noise-aware may differ substantially in the depth of its noise-related methodology.

\subsection{Internal Validity}

Internal validity concerns whether the distributions and relationships reported in the study reflect genuine properties of the analyzed literature rather than artifacts of the classification process.

\emph{Single-category assignment.} Each paper was assigned a single dominant testing category even though some contributions combine multiple testing paradigms. This threat is particularly relevant in QST because debugging, validation, assertion-based testing, and oracle construction frequently coexist within the same methodological workflow. We mitigated this threat by adopting a methodology-oriented assignment strategy that prioritizes the paper's primary contribution rather than all supporting mechanisms equally. The ambiguity-aware review process was specifically designed to identify and inspect such overlap cases.

\emph{Priority-hierarchy influence.} The deterministic priority hierarchy may influence the assignment of a small number of borderline papers, and alternative hierarchies could produce limited reassignment among ambiguous cases. All review-triggering conditions, reassignment decisions, and rationales are documented in the retained review log.

\emph{Source of evaluation-realism attributes.} For the 59 initial studies, the evaluation-realism and practice-pattern attributes were taken from the publicly released extraction of the source corpus (Section~\ref{subsec:evaluation_extraction}). Any extraction errors in that source would propagate to Sections~\ref{subsec:evaluation_realism} and~\ref{subsec:practice_patterns}; the conservative reporting rule limits, but does not remove, this dependence.

\emph{Residual category interpretation.} The residual \emph{Other} category may contain contributions that partially align with existing paradigms but do not satisfy the minimum lexical evidence threshold under the adopted operationalization. We intentionally retain this category as a conservative design choice to avoid forcing weakly supported assignments. Consequently, the category should not be interpreted purely as a collection of unrelated work or classification failures.

\subsection{External Validity}

External validity concerns the generalizability of the findings to the broader QST ecosystem and to future publication cohorts.

\emph{Dataset representativeness.} The corpus was assembled from major digital libraries with snowballing (Section~\ref{subsec:dataset_construction}) and captures much of the visible academic QST literature between 2018 and 2025, but it may not fully represent industrial practice, gray literature, workshop discussions not indexed by major repositories, or non-English publications.

\emph{Publication bias.} As with most mapping studies, the analyzed corpus necessarily reflects the subset of work that was successfully published or publicly disseminated. Methodologies that produced negative results, were abandoned during development, or did not satisfy publication expectations are not represented. The reported distributions should therefore be interpreted as characterizations of the published QST literature rather than of all QST methodologies explored in practice.

\emph{Small-sample categories.} Several testing paradigms contain relatively small numbers of papers (e.g., metamorphic testing and differential testing). Consequently, percentage-based comparisons involving these categories should be interpreted cautiously and regarded primarily as descriptive observations rather than statistically stable estimates.

\emph{Temporal completeness.} The study excludes partial-year publication data beyond 2025 to avoid distortions caused by incomplete indexing and publication cycles. As a result, subsequent methodological developments and recently emerging noise-aware frameworks are not reflected in the present analysis.

\subsection{Taxonomy Granularity and Robustness}
\label{subsec:threats_taxonomy}

A further validity concern is that treating Unit Testing and Coverage-Based Testing as independent paradigms may influence cross-category trends. The overall noise-awareness rate (23.0\%) is invariant under taxonomy restructuring because it is computed over the complete dataset. Table~\ref{tab:six_category_robustness} reports the explicit six-category aggregation in which Unit Testing and Coverage-Based Testing are subsumed under a broadened Test Generation paradigm. Three observations follow from the aggregated counts. First, the broadened Test Generation paradigm contains 25 papers, the largest merged paradigm in the corpus, with a noise-awareness rate of 20.0\%, still below the corpus mean of 23.0\%. The relative ordering of paradigms along the volume axis is therefore preserved, and the high-volume-low-awareness regime that Section~\ref{subsec:synthesis} identifies under the eight-paradigm taxonomy persists under aggregation. Second, no merged paradigm clears the 50\% majority threshold at $n>6$: the highest rate among paradigms with more than six papers remains that of Oracle/Assertion (27.3\%), while the residual \emph{Other} category reaches 50.0\% at $n=6$. Third, the merging hides genuine within-paradigm heterogeneity: the broadened Test Generation paradigm now combines a noise-awareness rate of 14.3\% from the test-generation subset, 50.0\% from the unit-testing subset, and 0.0\% from the coverage-based subset, producing an aggregate rate (20.0\%) that obscures both the comparatively hardware-aware unit-testing profile and the categorical absence of noise-aware coverage-based contributions.

\begin{table}[htbp]
\caption{Robustness analysis under a reduced six-category taxonomy formed by merging Unit Testing and Coverage-Based Testing into the Test Generation paradigm. The principal findings remain stable under aggregation: the merged Test Generation category remains below the corpus-wide noise-awareness rate, and no paradigm with more than six studies exceeds 50\% noise-awareness. $^\dagger$Merged category comprising Test Generation, Unit Testing, and Coverage-Based Testing.}
\label{tab:six_category_robustness}
\centering
\footnotesize
\begin{tabular}{lcc}
\toprule
\textbf{Aggregated paradigm}       & \textbf{N}  & \textbf{Noise-Aware (\%)} \\
\midrule
Test Generation (merged)$^\dagger$ & 25          & 20.0 \\
Oracle / Assertion                 & 11          & 27.3 \\
Debugging / Repair                 & 11          & 18.2 \\
Other                              & 6           & 50.0 \\
Mutation Testing                   & 4           & 25.0 \\
Metamorphic                        & 2           & 0.0 \\
Differential                       & 2           & 0.0 \\
\midrule
\textbf{Total}                     & \textbf{61} & \textbf{23.0} \\
\bottomrule
\end{tabular}
\end{table}

\subsection{Sensitivity to Adjudicated Labels}
\label{subsec:threats_sensitivity}

The corpus-level results use the framework's final labels rather than the adjudicated ground truth of the validation sample (Section~\ref{subsubsec:interrater_validation}). Re-running the analysis with the adjudicated labels substituted for all 31 validation papers leaves the principal findings unchanged. The number of noise-aware papers remains 14 (23.0\%). Test generation (1 of 13, 7.7\%) and debugging/repair (2 of 10, 20.0\%) remain below the corpus rate, oracle/assertion-based testing remains above it (5 of 13, 38.5\%), and coverage-based, metamorphic, and differential testing remain at zero. The paradigm-level association remains non-significant (permutation test of the paradigm$\times$noise-awareness table, $p \approx 0.34$; the same test gives $p \approx 0.39$ under the reported labels, in line with the Freeman--Halton $p \approx 0.40$ of Section~\ref{subsec:by_technique}), and the association between noise-awareness and hardware-realistic evaluation strengthens (odds ratio 115, $p < 0.001$). The temporal trend in the noise-aware share weakens (Cochran--Armitage $p \approx 0.26$), reinforcing the descriptive reading of RQ3.

\subsection{Conclusion Validity}

Conclusion validity concerns the appropriateness of the interpretations derived from the empirical findings. The study is primarily descriptive rather than inferential in nature; accordingly, the reported percentages and distributions should be interpreted as characterizations of the analyzed literature rather than as predictive estimates of future research trajectories.

Several findings additionally depend on relatively small subsets of papers. For example, the comparatively high noise-awareness rate observed in Unit Testing is based on only six papers, and the paradigm-level association is not statistically significant (Section~\ref{subsec:by_technique}). Such findings are methodologically suggestive but should not be overgeneralized.

The temporal trends reported in the study should likewise be interpreted conservatively. Annual publication cohorts are relatively small, and fluctuations between years may reflect short-term publication dynamics rather than stable longitudinal behavior. The broader multi-year pattern is more robust than individual year-to-year changes, but the Cochran--Armitage test (Section~\ref{subsec:temporal_noise}) shows that the increase in the noise-aware share falls just short of statistical significance ($p \approx 0.051$) and that the hardware-realistic share shows no significant trend; temporal claims are therefore stated descriptively.

Finally, the study does not claim that noise-aware methodologies are necessarily superior to noise-oblivious methodologies in all contexts. Some testing paradigms may legitimately focus on idealized reasoning, algorithmic validation, or theoretical analysis rather than on direct hardware-oriented evaluation. The analysis instead highlights the relative extent to which contemporary QST methodologies explicitly engage with noisy and hardware-constrained execution.

\subsection{Reliability}

The automated component is fully reproducible: the same lexicons, thresholds, hierarchy, ambiguity rules, and decision logic (Algorithm~\ref{alg:classification}) applied to the same metadata yield identical outputs, and classification logs, flags, and manual-review decisions with written rationale are retained so that reassignments can be inspected and replicated. The manual-review stage introduces limited expert interpretation; the blinded inter-rater study (Section~\ref{subsubsec:interrater_validation}) quantifies this directly ($\kappa = 0.962$ for testing technique and $\kappa = 0.760$ for noise-awareness), and manual inspection is confined to flagged papers. Where papers did not explicitly report evaluation or hardware-related details, we avoided inferring unsupported backend behavior.

\section{Conclusion}
\label{sec:conclusion}

This paper presented a systematic empirical landscape analysis of noise-awareness in quantum software testing (QST) research through a deterministic taxonomy-guided classification framework applied to 61 papers published between 2018 and 2025. The study combined automated lexical classification, ambiguity-aware review mechanisms, targeted manual inspection, and paradigm-level synthesis to characterize how contemporary QST methodologies engage with noisy and hardware-constrained quantum execution.

The findings consistently indicate that noise-awareness remains a minority concern across the current QST landscape. Among the analyzed papers, 77.0\% were classified as noise-oblivious, while only 23.0\% explicitly incorporated noise-related execution concerns under the operational criteria adopted in this study. Although recent years show gradual growth in hardware-conscious testing research, the dominant paradigms in the field, most notably test generation and debugging/repair, remain primarily grounded in idealized execution assumptions, while coverage-based, metamorphic, and differential testing contain no noise-aware contributions in the analyzed corpus.

Beyond individual paradigm statistics, the results reveal a broader structural property of the QST ecosystem: no testing paradigm currently combines both high publication volume and consistently strong noise-awareness. The methodological region corresponding to simultaneously mature and hardware-conscious QST practice therefore remains largely unoccupied. While recent frameworks such as HornBro, QUT, and Bloch Vector Assertions illustrate growing interest in hardware-aware testing and validation, these contributions currently appear as isolated methodological efforts rather than as components of mature and widely adopted research subfields.

The study additionally highlights methodological gaps in differential, coverage-based, and metamorphic testing, and, more broadly, the absence of widely reusable mechanisms for treating uncertainty, statistical confidence, and hardware-noise characteristics as first-class testing concerns (Section~\ref{subsec:disc_implications}).

A third reading of the corpus, through the empirical-practice patterns aggregated in Section~\ref{subsec:practice_patterns}, adds a complementary diagnostic. The methodological elements most directly required by probabilistic NISQ execution are also the least consistently documented: explicit test oracles and shot budgets are reported in little more than half of the papers, and inferential statistical tests in only 16.4\%. Noise-aware papers document oracles and shot budgets more often than noise-oblivious papers, but inferential statistical tests are reported at similarly low rates in both subsets, indicating that the field's transition to hardware-conscious testing will require methodological development at the reporting-practice layer as well as at the testing-methodology layer.

Methodologically, the study contributes more than a descriptive taxonomy of QST paradigms. The proposed framework explicitly addresses semantic overlap, ambiguity, and cross-category contamination within QST literature through priority-based tie-breaking, ambiguity-aware review flagging, semantic inspection, and structured manual-review integration. The results demonstrate that purely lexical classification is insufficient for several classes of QST papers, particularly those combining oracle-oriented terminology with debugging-oriented methodological contributions. The framework's empirical reliability is supported on two levels. Internally, the deterministic pipeline reproduces 56 of 61 final classifications (91.8\%), with every pipeline--final disagreement surfaced by the flag system itself. Externally, a blinded inter-rater study on 31 papers yielded almost perfect agreement on testing-technique classification ($\kappa = 0.962$) and substantial agreement on noise-awareness ($\kappa = 0.760$), and the classifier reproduced the adjudicated human ground truth at substantial-to-almost-perfect levels. In addition, the abstract-level noise label is strongly associated with the evaluation backends extracted from the full papers (odds ratio 40.5, $p < 0.001$), and the corpus-level findings are unchanged when the adjudicated labels are substituted. The resulting rule-based-yet-reviewable workflow therefore constitutes an independently useful contribution for future empirical studies of QST and adjacent quantum software engineering domains.

The findings further suggest that the long-term maturation of QST will likely require a broader methodological transition from simulator-centric validation toward systematically hardware-conscious testing design. Achieving such a transition will require not only new testing techniques, but also reusable noise-aware oracle mechanisms, probabilistic adequacy criteria, cross-platform validation infrastructures, and empirical benchmarking practices capable of supporting realistic NISQ-era evaluation.

Finally, the analytical framework introduced in this study remains reusable beyond the present dataset. The deterministic classification pipeline, ambiguity-aware review methodology, noise-awareness operationalization, and longitudinal analysis strategy may be extended to future publication cohorts and to adjacent QSE subareas such as quantum verification, quantum benchmarking, and quantum continuous integration. Maintaining such longitudinal empirical monitoring may help the research community systematically track the evolution of hardware-conscious testing practice as the broader quantum software ecosystem matures.

\section*{Declarations}

\noindent\textbf{Competing interests:} The authors declare no competing interests.

\noindent\textbf{Data availability:} The classification data, keyword lexicons, and manual-review log are available from the corresponding author upon request.

\bibliographystyle{IEEEtran}
\bibliography{references}

\end{document}